\documentclass[10pt,letterpaper]{article}
\usepackage[top=0.85in,left=2cm, right=5cm,footskip=0.75in]{geometry}
\usepackage{changepage}

\usepackage[group-separator={,}]{siunitx}

\usepackage[utf8]{inputenc}

\usepackage{color,textcomp,marvosym}

\usepackage{fixltx2e}

\usepackage{amsmath,amssymb,amsthm}

\usepackage{cite}

\usepackage{nameref,hyperref}

\usepackage[right]{lineno}

\usepackage{microtype}
\DisableLigatures[f]{encoding = *, family = * }

\usepackage{rotating}

\usepackage{graphicx}

\newcommand{\defeq}{\overset{\text{def}}{=}}

\newtheorem*{rem}{Remark}

\date{}

\usepackage{lastpage,fancyhdr,graphicx}
\usepackage{epstopdf}
\begin{document}
\vspace*{0.35in}

\begin{flushleft}
{\Large
\textbf\newline{Somatic Mutations Render Human Exome and Pathogen DNA more Similar}
}
\newline
\\
Ehsan Ebrahimzadeh\textsuperscript{1},
Maggie Engler\textsuperscript{2},
David Tse\textsuperscript{2},
Razvan Cristescu\textsuperscript{3},
Aslan Tchamkerten\textsuperscript{4,* }
\\
\bigskip
\bf{1} Department of Electrical Engineering, UCLA, Los Angeles, California, USA
\\
\bf{2} Department of Electrical Engineering, Stanford University, Stanford, California, USA
\\
\bf{3} Department of Discovery Medicine, Merck Research Laboratories, Rahway, New Jersey, USA
\\
\bf{4} Department of Communications and Electronics, Telecom ParisTech, Paris, France
\bigskip

%
%





* aslan.tchamkerten@telecom-paristech.fr
\end{flushleft}
\section*{Abstract}
Immunotherapy has recently shown important clinical successes in a substantial number of oncology indications.  Additionally, the tumor somatic mutation load has been shown to associate with response to these therapeutic agents, and specific mutational signatures are hypothesized to improve this association, including signatures related to pathogen insults. We sought to study in silico the validity of these observations and how they relate to each other. We first addressed the question whether somatic mutations typically involved in cancer may increase, in a statistically meaningful manner, the similarity between common pathogens and the human exome. Our study shows that common mutagenic processes like those resulting from exposure to ultraviolet light (in melanoma) or smoking (in lung cancer) increase,  in the upper range of biologically plausible frequencies, the similarity between cancer exomes and pathogen DNA at a scale of $12$ to $16$ nucleotide sequences (corresponding to peptides of $4-5$ amino acids). Second, we investigated whether this increased similarity is due to the specific mutation distribution of the considered mutagenic processes or whether uniformly random mutations at equal rate would trigger the same effect. Our results show that, depending on the combination of pathogen and mutagenic process, these effects need not be distinguishable. Third, we studied the impact of mutation rate and showed that increasing mutation rate generally results in an increased similarity between the cancer exome and pathogen DNA, again at a scale of $4-5$ amino acids.  Finally, we investigated whether the considered mutational processes result in amino-acid changes with functional relevance that are more likely to be immunogenic.  We showed that functional tolerance to mutagenic processes across species generally suggests more resilience to mutagenic processes that are due to exposure to elements of nature than to mutagenic processes that are due to exposure to cancer-causing artificial substances. These results support the idea that recognition of pathogen sequences as well as differential functional tolerance to mutagenic processes may play an important role in the immune recognition process involved in tumor infiltration by lymphocytes.



\section*{Introduction}
Recent clinical advances firmly establish the role of immunotherapy (in particular, checkpoint inhibition targetting the CTLA4 and PD1/PD-L1 pathways \cite{hoos}) in the treatment of cancer. However, the rates of response vary by indication, outlining the important role of identifying the patients most likely to respond \cite{mariathasan2018tgfbeta,chowell2018patient,riaz2017tumor,luksza2017neoantigen}. In parallel, the analysis of the data in large scale genomic efforts including The Cancer Genome Atlas (TCGA \cite{TCGA})  has identified universal characteristics of the tumor and its environment that ellicit potential recognition by the host immune system.  In particular, somatic mutational load as inferred by DNA sequencing \cite{alexandrov2013signatures,lawrence2013mutational} and cytolytic infiltrate as inferred by immunohistochemistry or RNA sequencing \cite{rooney2015molecular} have emerged as hallmarks of an immune-active tumor enviroment.   It is thus important to understand the causality and mechanism of action that drives the heterogenous composition of the tumor and its environment and consequently the heterogeneity of response to immunotherapy, in order to select the right patients for treatment, potential combinations, and potential for early intervention.  

Multiple recent studies have suggested a strong causal link between the mutational burden of the tumor and clinical response to immunotherapy across multiple indications including Melanoma \cite{van2015genomic,snyder2014genetic}, Non Small Cell Lung Cancer \cite{rizvi2015mutational}, Bladder cancer \cite{rosenberg2016atezo} and Colorectal cancer \cite{le2015pd}.  In these studies, a strong relationship between neoantigen load (the number of mutations with immunogenic potential) and response to immunotherapy has been identified.  Importantly, each of these indications are characterized by distinct mutagenic processes that result in abundant neoantigen load \cite{alexandrov2013signatures,lawrence2013mutational}: UV light exposure in Melanoma, smoking in Non Small Cell Lung Cancer, APOBEC activation in Bladder cancer, and MMR defficiency in MSI-h Colorectal cancer.  Whether particular mutations or mutational patterns preferentially induce an immunologic phenotype remains an open question \cite{van2015genomic,snyder2014genetic}.  However, several hypotheses have recently been put forward, including the presence of mutations in particular genes \cite{hugo2016genomic,eroglu2018high}, or the presence of a transversion signature related to smoking \cite{rizvi2015mutational}. In particular, Snyder \emph{et al.} \cite{snyder2014genetic} put forward a hypothesis linking cancer exomes with patterns present in common pathogens.  Namely, their results with exome analysis of Melanoma patients treated with Ipilimumab, a CTLA4 inhibitor, suggest that somatic mutations in cancer genomes that lead to tetrapeptides similar to those found in common pathogens are more likely to elicit a response to the therapy than common somatic mutations. This association is presumably driven by the innate ability of significant portions of the adaptive immune repertoire to recognize such pathogens.

We took an in-silico approach to evaluate the impact of certain mutagenic processes on the similarity between cancer exomes and pathogen DNAs. Somatic mutations are an inherent natural process related to cell division and aging which in some instances is exacerbated by mutagenic factors.  We simulated such mutagenic processes using mixtures of mutational signatures with empirically derived mixing parameters.  We used a simple similarity metric between the mutated exome and common pathogen exomes to estimate changes in overall potential immunogenicity of cancer exomes as compared to the normal exome. We considered simulations of mutagenic processes that yield most mutated cancer exomes, namely ultra-violet (UV) light (Melanoma), smoking (Non Small Cell Lung Cancer), and APOBEC activation (Bladder cancer) \cite{rooney2015molecular,alexandrov2013signatures}.  Our results suggest that, in the upper range of biologically plausible mutation rates, mutagenic processes resulting from exposure to these common mutagens lead to cancer exomes that are more similar pathogen DNAs at a scale of $12$ to $16$ nucleotides. These changes are subtle but nevertheless statistically significant and are particularly important in the range of peptide sizes  that are relevant for epitope presentation in the 
 human MHC mechanism; MHC presentation typically involves peptides with lengths between 8-18 nucleotides (8-13 for class I MHC and 13-18 for class II MHC \cite{wieczorek2017major}).

However, our results also suggest that the increased similarity need not be caused by the specificity of the mutation distribution. Depending on the pathogen, uniformly random mutations (at the same rate) may result in equal increased similarity. Finally, we show that increasing mutation rate generally results in increased similarity between cancer exomes and pathogen DNAs. These conclusions suggest that mutagenic processes might act as a mechanism of pressure that models the mutational spectra observed in tumors by increasing recognition from the host immune system. 

Opposite to the aforementioned effect that increases the likelihood that a cancer exome is recognized by the immune system, an antagonist mechanism of pressure on mutational landscape stems from tolerance by the immune system to natural mutagenic processes. To that extent, we establish that exomes across species are generally more resilient, in terms of a functional point of view related to the synonymity of amino-acid changes,  to  mutagenic processes that are due to exposure to elements of nature than to mutagenic processes that are due to exposure to cancer-causing artificial substances. 
 In particular, we observe that the functionality of the genetic code (allocation of codons to amino-acids) is more resilient to UV light than smoking  mutagenic processes at a fixed rate.  This suggests the possibility that there are  different tissue-dependent evolutionary tolerance levels, modulated by the pathogen recognition apparatus in terms of both immune recognition and cancer development, which for example reflect in the much higher mutational loads and immune infiltrate in Melanoma compared to Lung cancer \cite{rooney2015molecular}.  

\section{Methods}
We sought to assess whether certain mutagenic processes result in somatic alterations that increase the similarity of the mutated human exome with selected pathogens. Accordingly, we first defined a pairwise similarity metric among DNA sequences of different length and evaluated the similarity between pathogens and the normal human exome.  Second, we simulated mutations resulting from different mutagenic processes at different mutation rates acting on the human exome and evaluated the consequent change in similarity of the mutated human exome with respect to the pathogen exomes. Third,  we investigated the resiliency of exomes (human exome and model organism exomes) in terms of maintained functionality of the resulting amino-acids and compared the sequences of amino acids of the normal and mutated exomes.
\subsection*{Data and computing resources}
We obtained the human normal exome from GRCh38 \url{http://www.ensembl.org/Homo_sapiens/Info/Index} 

We considered the following list of model organisms: Mus Musculus (Mouse), Saccharomyces Cerevisiae (Yeast), Felis Catus (Cat), Drosophila Melanogaster (Fruitfly),  Caenorhabditis Elegans (Nematode), Xenopus,  Danio Rerio (Zebrafish), Cavia Porcellus (Pig), Anolis carolinensis (Anolis). Exomes from these organisms were obtained from 
\url{http://uswest.ensembl.org/biomart/martview/}

We considered the following list of viral pathogens: Cytomegalovirus (CMV), Dengue virus, Ebola virus, Epstein-Barr virus (EBV), Human Herpesvirus 6 (HHV), Human Papillomavirus  (HPV), Measles virus, Yellow Fever virus. DNA sequences from these pathogens were obtained from \url{http://www.ncbi.nlm.nih.gov/}

We considered simulations of mutational signatures resulting from ultra-violet (UV) light (specific to Melanoma), smoking (specific to Non Small Cell Lung Cancer (NSCLC)), and APOBEC activation (specific to Bladder cancer). These simulations were based on the data from \cite[Supplementary information, Table S2]{lawrence2013mutational} restricted to the set of patients with Melanoma cancer, NSCLC, and Bladder cancer.

For simulations we used Python 2.7.6 (libraries random, numpy, and scipy.stats) and ran programs on a shared server with 8 CPUs and 128GB memory.

\section{Results}

\subsection{Pathogen DNA vs. human exome and MHC mechanism  }\label{res_pat}

To quantify the similarity between a pathogen DNA, denoted by $x$, and the human exome, denoted by $y$, we considered the following similarity score. For a given integer $\ell \geq 1$,  the similarity score, denoted by $s_\ell(x,y)$, corresponds to the relative proportion of length-$\ell$ strings in the pathogen DNA that also appear in the human exome at least once, that is
$$s_\ell(x,y)\defeq \frac{1}{L-\ell +1}\sum_{i=1}^{L-\ell+1}z_i$$
where $$z_i\defeq \left\{ \begin{array}{ll} 1 & \text{if}\quad  x_{i}^{i+\ell-1}\prec y\\
0 & \text{otherwise.} \end{array}\right.$$
Here $L$ denotes the length of the pathogen DNA, $x_{i}^{i+\ell-1}\defeq x_i,x_{i+1},\ldots,x_{i+\ell-1}$ denotes the pathogen DNA substring starting at position $i$ and ending at position $i+\ell-1$, and ``$\prec$'' denotes string inclusion.
 In particular, $s_\ell(x,y)=1$ corresponds to the case where all length-$\ell$ strings in the pathogen DNA also appear in the human exome and  $s_\ell(x,y)=0$ corresponds to the case where the pathogen DNA and the human exome have no length-$\ell$ string in common. Observe that $s_\ell(x,y)$ can be interpreted as the probability that a randomly and uniformly picked length-$\ell$ string in the pathogen DNA also appears in the human exome. Accordingly, we often refer to $s_\ell(x,y)$ as the matching probability. Finally, notice that  $s_\ell(x,y)$ does not count multiplicity, {\it{i.e.}}, strings that appear only once in the human exome and strings that appear multiple times in the human exome are note distinguished

\begin{figure}
    \centering
    \includegraphics[scale=.5]{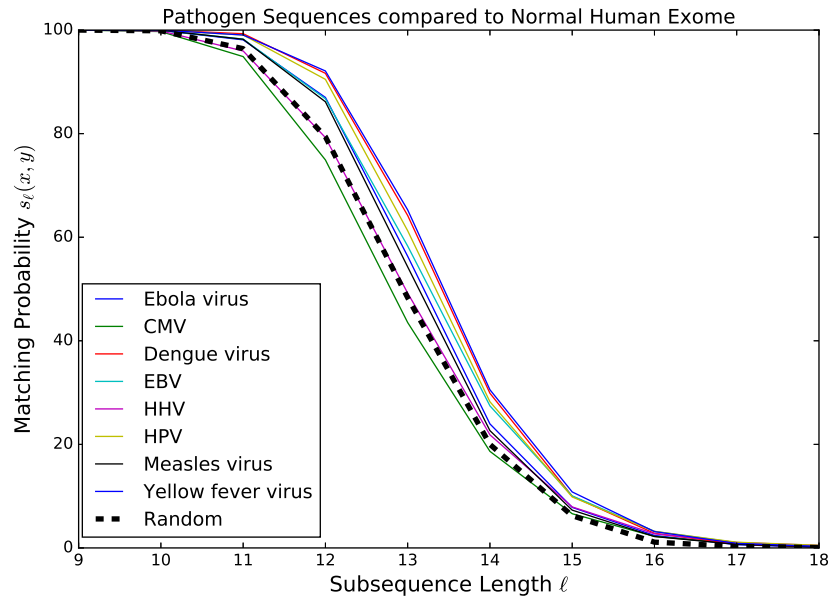}
    \vspace{.2cm}
    \caption{Each curve represents the matching probability (similarity score) $s_\ell(x,y)$ between a pathogen DNA  $x$ and the human exome $y$, as a function of the  subsequence length $\ell$. The ``Random'' curve refers to the average score of a randomly and uniformly generated ``pathogen'' DNA sequence. }
    \label{fig:plot1}
\end{figure}
In Fig.~\ref{fig:plot1}, each curve represents the matching probability $s_\ell(x,y)$ for a specific pathogen DNA $x$ and the normal human exome $y$, for $\ell\in \{9,10,\ldots,18\}$. To benchmark these scores we also considered the matching probability with respect to a randomly and uniformly generated ``pathogen'' sequence, where each nucleotide is equally likely to occur. The average matching probability with respect to such a sequence is represented by the ``Random'' curve in Fig.~\ref{fig:plot1} and turns out to be independent of its length $L$. This curve is indistinguishable from the $95\%$ confidence interval corresponding to a randomly generated sequence. Supporting material for Fig.~\ref{fig:plot1} is deferred to Section~\ref{supportaa} in the Appendix. We make the following observations:
\begin{itemize}
\item[$\diamond$]
For all pathogens the similarity score is equal to one for $\ell\leq 10$, that is length $\ell\leq 10$ subsequences of the pathogen DNAs all appear in the human exome as well.
\item[$\diamond$] The similarity scores are non-zero for all pathogens up to length $\ell=20$. At $\ell=21$ the similarity scores is zero for  the Ebola virus, the Measles virus, and the Dengue virus.
\item[$\diamond$]
For all $\ell\in  \{11,\ldots,18\}$ the similarity score for pathogen DNAs is higher than for a random sequence,  except for CMV ($\ell\in \{15,\ldots,18\}$) and for HHV ($\ell\in \{13,\ldots,18\}$).
\item[$\diamond$] From $\ell=10$ there is a steep decrease in the similarity scores, down to less than $15\%$ for $\ell=15$. A closer look at the data (see Table~\ref{matchoscore}) reveals that, for all pathogens, the sharpest relative drop of the similarity score occurs from $\ell=12$ to $\ell=13$ or from $\ell=13$ to $\ell=14$.
\item[$\diamond$] The differences in score across pathogens is maximal at $\ell\in \{12,13\}$.  
\end{itemize}
These in-silico observations are in line with the concept that $4-5$ amino acids are enough for the presentation machinery in terms of both diversity of possible sequences ($20^4-20^{5}$) and differentiation of self from foreign sequences  in the MHC machinery. Namely, this length is strikingly similar to the length of peptides studied in the signature determined by \cite{snyder2014genetic}.

\subsection{Impact of somatic mutations on pathogen DNA and human exome similarity score}

To assess the impact of somatic mutations on pathogen DNA and human exome similarity score and identify the roles of mutation distribution and mutation rate we proceeded as follows:
\begin{itemize}
\item
Normal exome vs. cancer exome: we investigated whether cancer somatic mutations render pathogen and human exome more similar, and whether random mutations alone, with uniform distribution across mutations, would produce the same results as (typically non-uniform) cancer-dependent mutations, at the same mutation rate.
\item Impact of mutation rate: we investigated whether a higher mutation rate renders pathogen DNA and human exome more similar.
\end{itemize}
Central to our investigation is a notion of cancer channel described next.
\subsubsection*{Cancer channel}
We  simulated the changes induced to the normal exome by cancer specific mutagens in a probabilistic way. The cancer exomes were generated from the normal exome by using cancer-dependent mixtures of mutational signatures with empirical weights derived from data in \cite{lawrence2013mutational}.  Note that even if a cancer typically exhibits a dominant mutational signature, the simulated mutagenic process results in a more realistic combination of such signatures. The similarity scores of the normal exome and cancer exome were then computed for each pathogen.
To formalize our analysis, we used concepts from information theory, in particular related to communications over a noisy channel. To a given cancer and mutation rate we associated a transformation, referred to as ``cancer channel,'' which mimics the typical effects of the mutagenic process that are specific to the cancer at the given mutation rate. Analogously to a communication channel that alterates a transmitted message because of noise (see, {\emph{e.g.}},\cite{cover2012elements}), a cancer channel alterates a DNA sequence because of somatic mutations. Given a particular cancer $c$ and a mutation rate $\rho$ the cancer channel assigns to each nucleotide $\alpha$ the probability ${\mathbb{P}}_{c,\rho}(\beta|\alpha)$ of being mutated into nucleotide $\beta$. This probability was derived using data from \cite[Supplementary information, Table S2]{lawrence2013mutational} (see Appendix~\ref{appB} in this paper). 

To obtain a cancer exome $\tilde{y}$ we  ``passed'' the normal human exome $y$ through cancer channel ${\mathbb{P}}_{c,\rho}(\cdot |\cdot)$ as shown in Fig.~\ref{channel}.  
\begin{figure}
\begin{center}
\caption{\label{channel}Effects of cancer specific mutations on a normal exome modeled as a cancer channel. Cancer exome $\tilde{y}=\{\tilde{y}_1,\tilde{y}_2,\ldots,\tilde{y}_G\}$ is obtained from normal exome $ y=\{y_1,y_2,\ldots,{y}_G\}$ through a cancer specific probabilistic transformation ${\mathbb{P}}_{c,\rho}(\beta|\alpha)$ which assigns to each nucleotide $\alpha$ the probability of being mutated to nucleotide $\beta$. This transformation depends on both the mutation distribution specific to cancer $c$ and the mutation rate $\rho$.}
\end{center}
\end{figure}
Specifically, the cancer exome $\tilde{y}$ was generated from ${y}$ so that the probability to obtain $\tilde{y}=\{\tilde{y}_1,\tilde{y}_2,\ldots,\tilde{y}_G\}$ from normal exome $y=\{y_1,y_2,\ldots,{y}_G\}$
was given by
$$\prod_{i= 1}^G{\mathbb{P}}_{c,\rho}(\tilde{y}_i|y_i).$$ 

\subsubsection*{Normal vs. cancer specific and random mutations}
For given pathogen $x$, cancer $c$, and mutation rate $\rho$ we performed two tests. In Test $1$, we evaluated the statistical significance of the effect of cancer somatic mutations in making human exome more similar to pathogen DNA sequences. In Test $2$, we compared cancer somatic mutations and random mutations in making the human exome more similar to pathogen DNA sequences. Both tests were peformed for  $\rho$-values of $0.0005$, $0.001$, and $0.01$. The lowest mutation rate was chosen to be $0.0005$ as it represents a good compromise between biological and statistical relevance. It lies in the upper range of the mutation rates observed in actual cancer samples \cite{lawrence2013mutational} and in the lower range for statistical relevance---see next subsection.

\noindent {{\textbf{Test 1:}}} For each $\ell \in \{10,11, \ldots,18\}$ we independently generated $1000$ cancer exomes $\{\tilde{y}\}$ from the normal human exome $y$ and computed the corresponding similarity scores $\{s_\ell(x,\tilde{y})\}$.  $P$-values were computed for comparing the mean of $\{s_\ell(x,\tilde{y})\}$ against $s_{\ell}(x,y)$ using a one-sided t-test with a null hypothesis that the true mean of $s_\ell(x,\tilde{y})$ is no larger than $s_{\ell}(x,y)$.

\noindent {{\textbf{Test 2:}}}  We replaced the cancer channel by  a ``random channel'' which produced mutations at the same rate but in a uniform $(1/3,1/3,1/3)$ manner. 
For each $\ell \in \{10,11, \ldots,18\}$ we independently generated $1000$ exomes $\{\hat{y}\}$ by passing the normal human exome $y$ through the random channel and computed the corresponding similarity scores $\{s_\ell(x,\hat{y})\}$. $P$-values were computed for comparing the mean of $\{s_\ell(x,\hat{y})\}$ against the mean of $\{s_\ell(x,\tilde{y})\}$ (obtained in Test~$1$) using a two-sample one-sided t-test with a null hypothesis that the true mean of $s_\ell(x,\tilde{y})$ is no larger than the true mean of $s_\ell(x,\hat{y})$---note that directly computing the true mean of $s_\ell(x,\tilde{y})$ over $\tilde{y}$ is impossible as it amounts to computing a sum over all $\approx 4^{10^7}$ possible cancer exomes, and similarly for the mean of $s_\ell(x,\hat{y})$.

\begin{figure}
    \centering
    \includegraphics[width=0.32\textwidth]{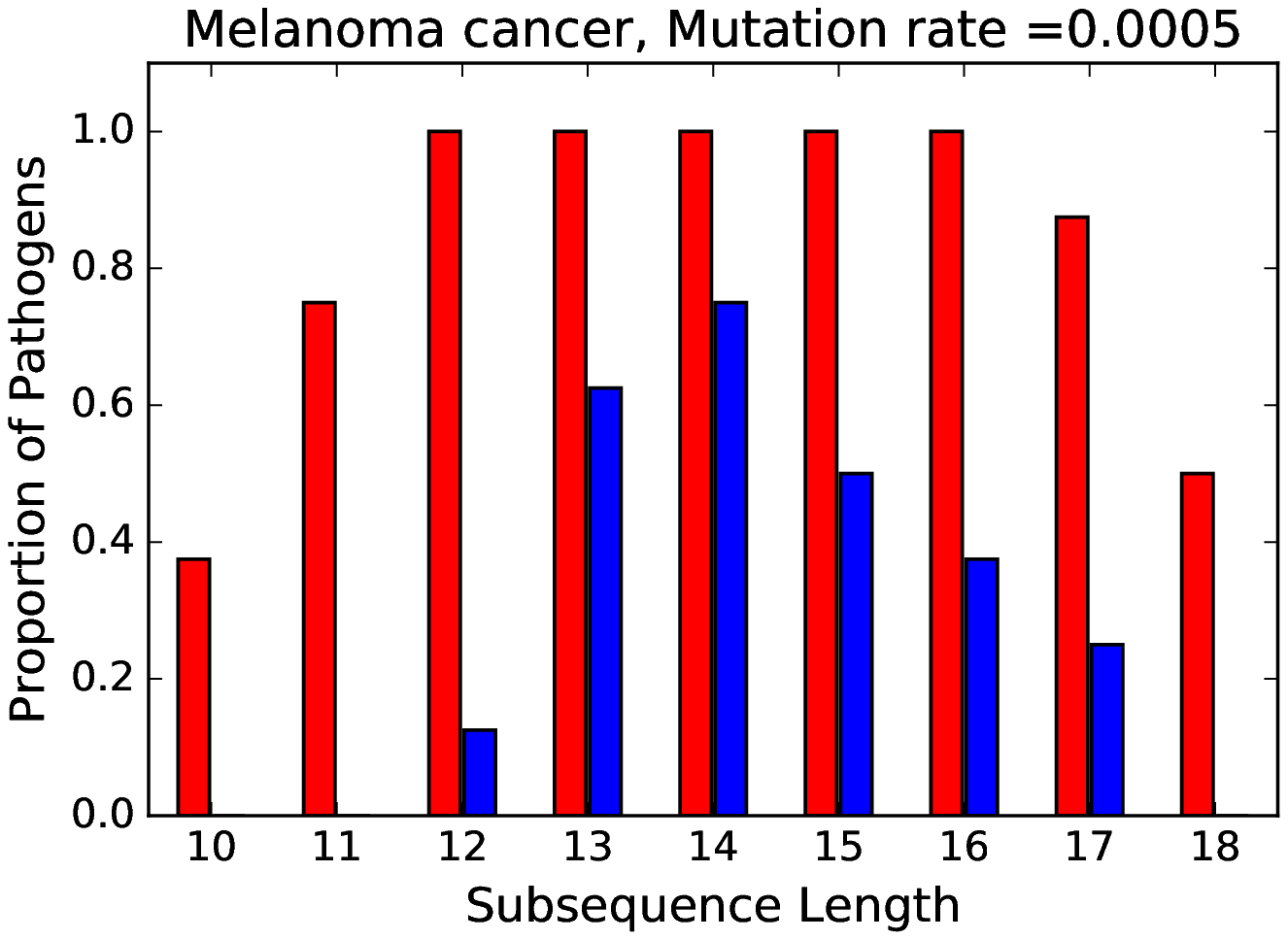}
\includegraphics [width=0.32\textwidth]{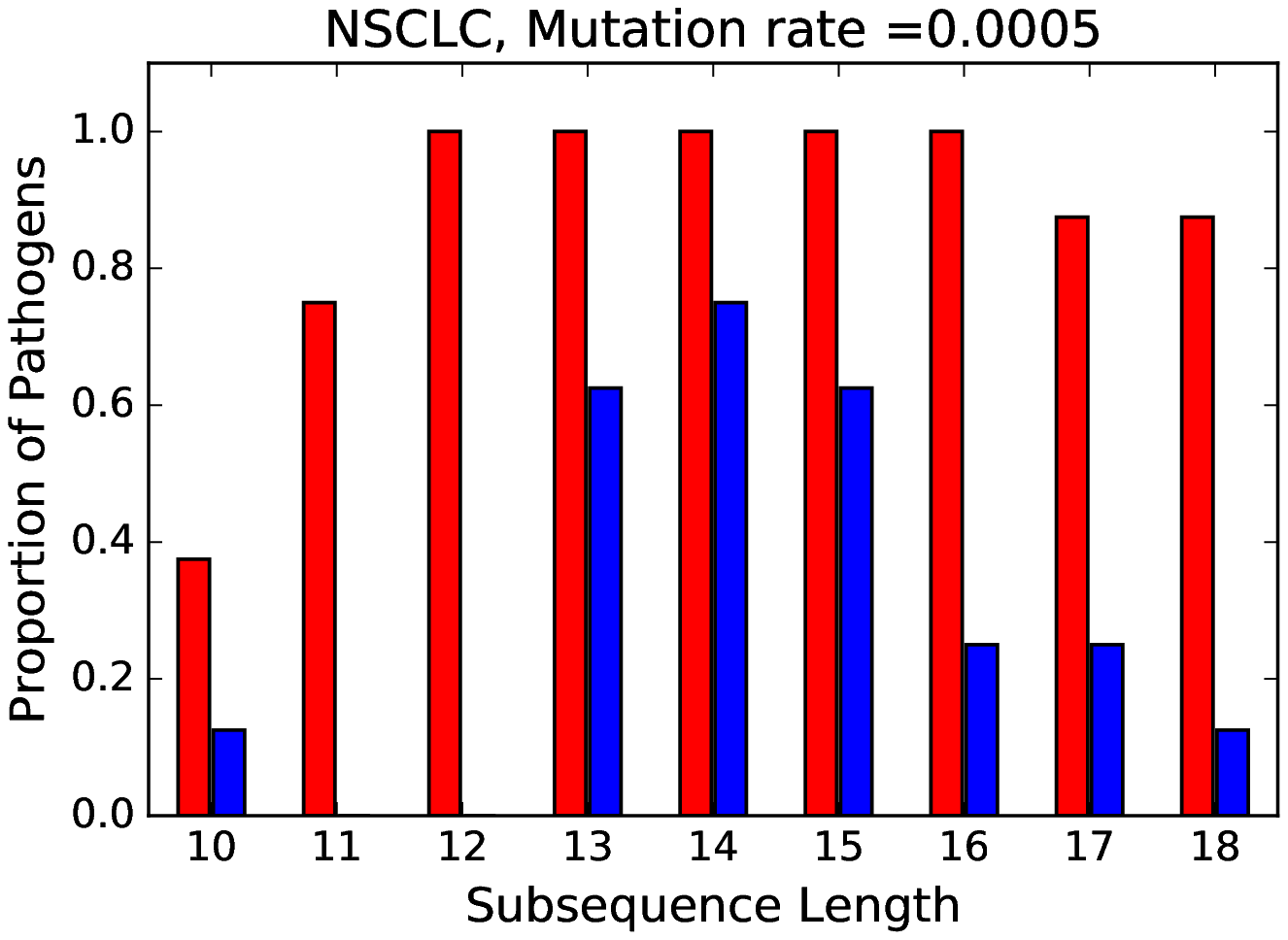}
    \includegraphics [width=0.32\textwidth]{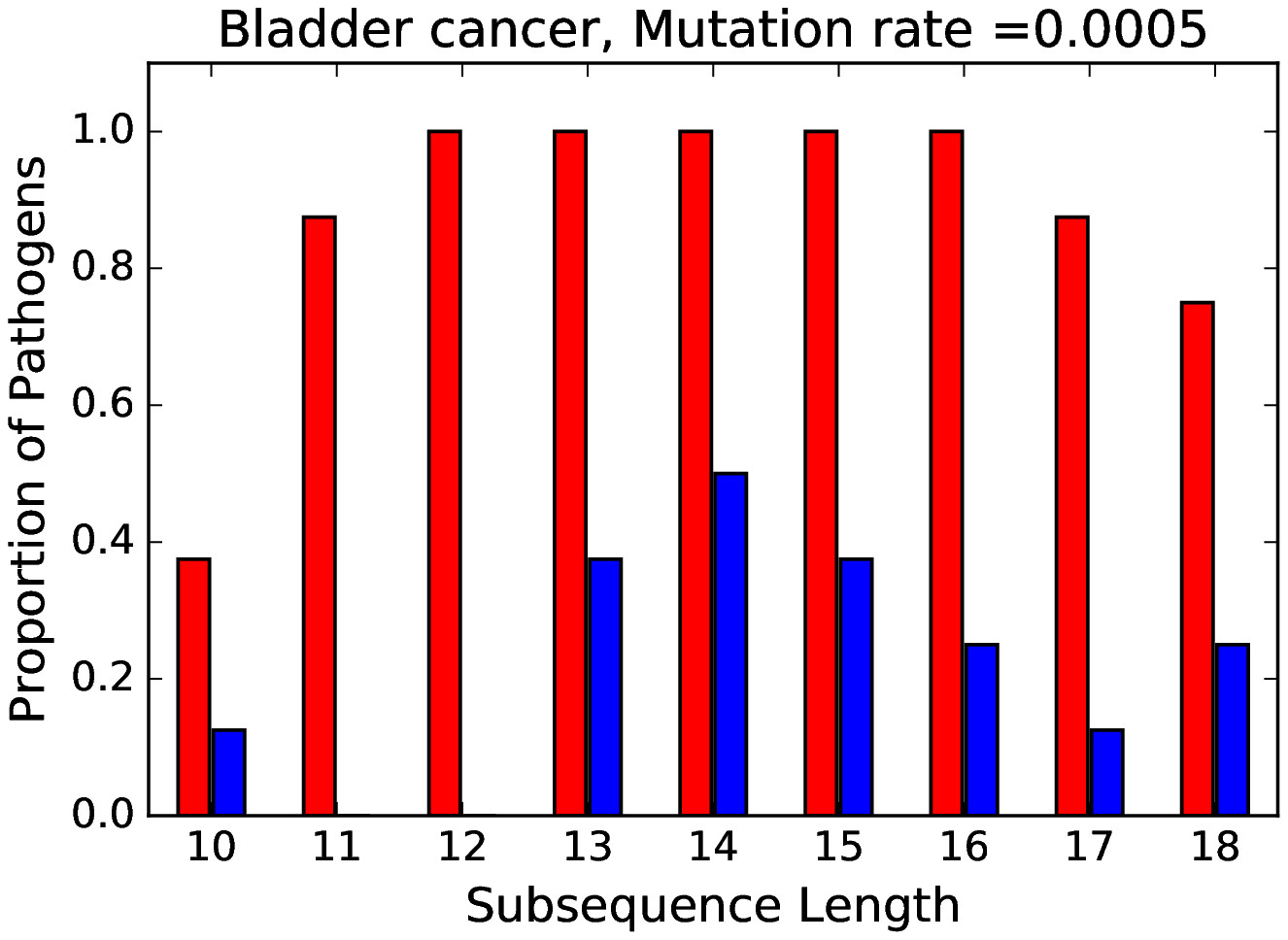}
   \includegraphics[width=0.32\textwidth]{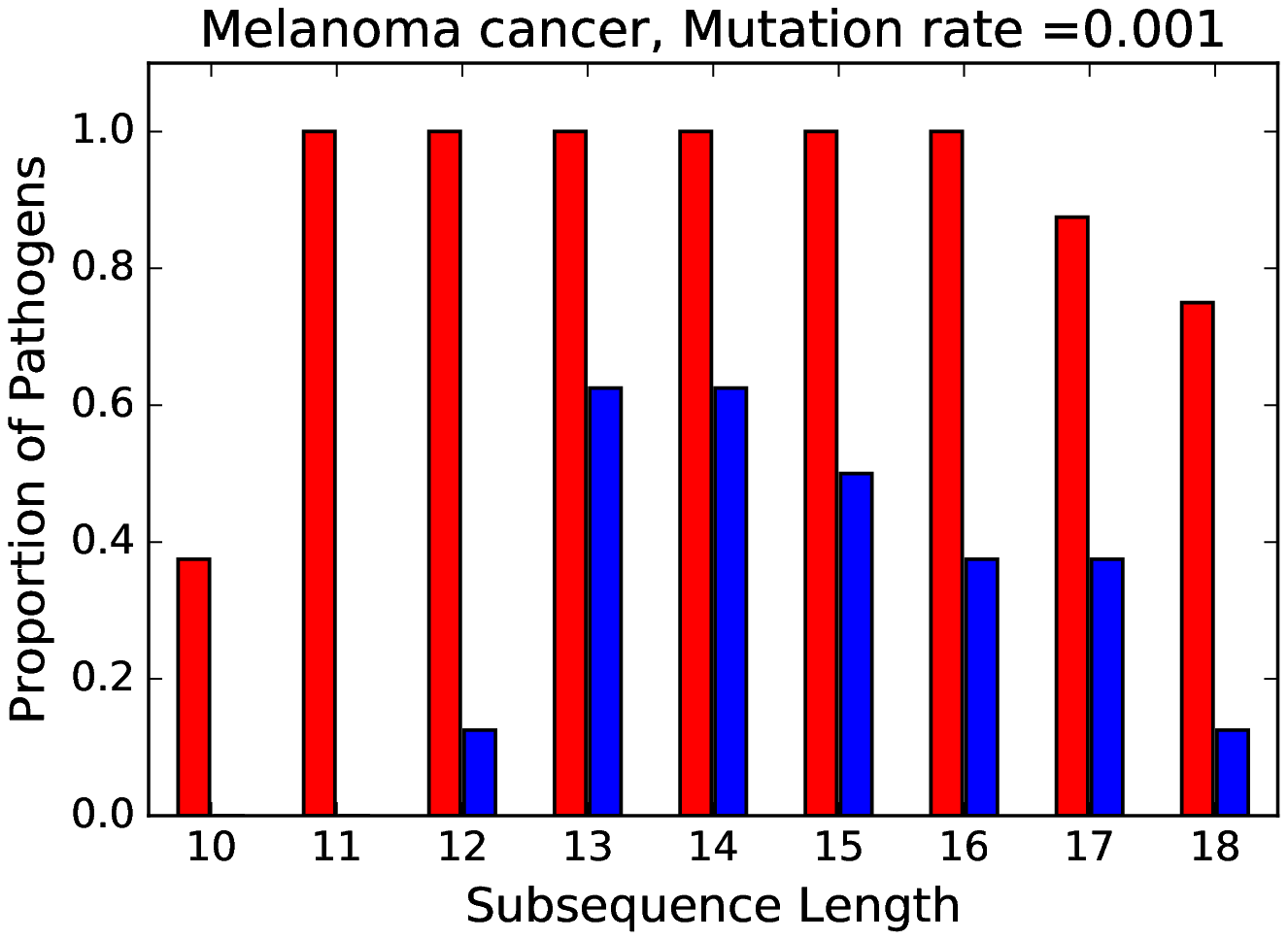}
         \includegraphics [width=0.32\textwidth]{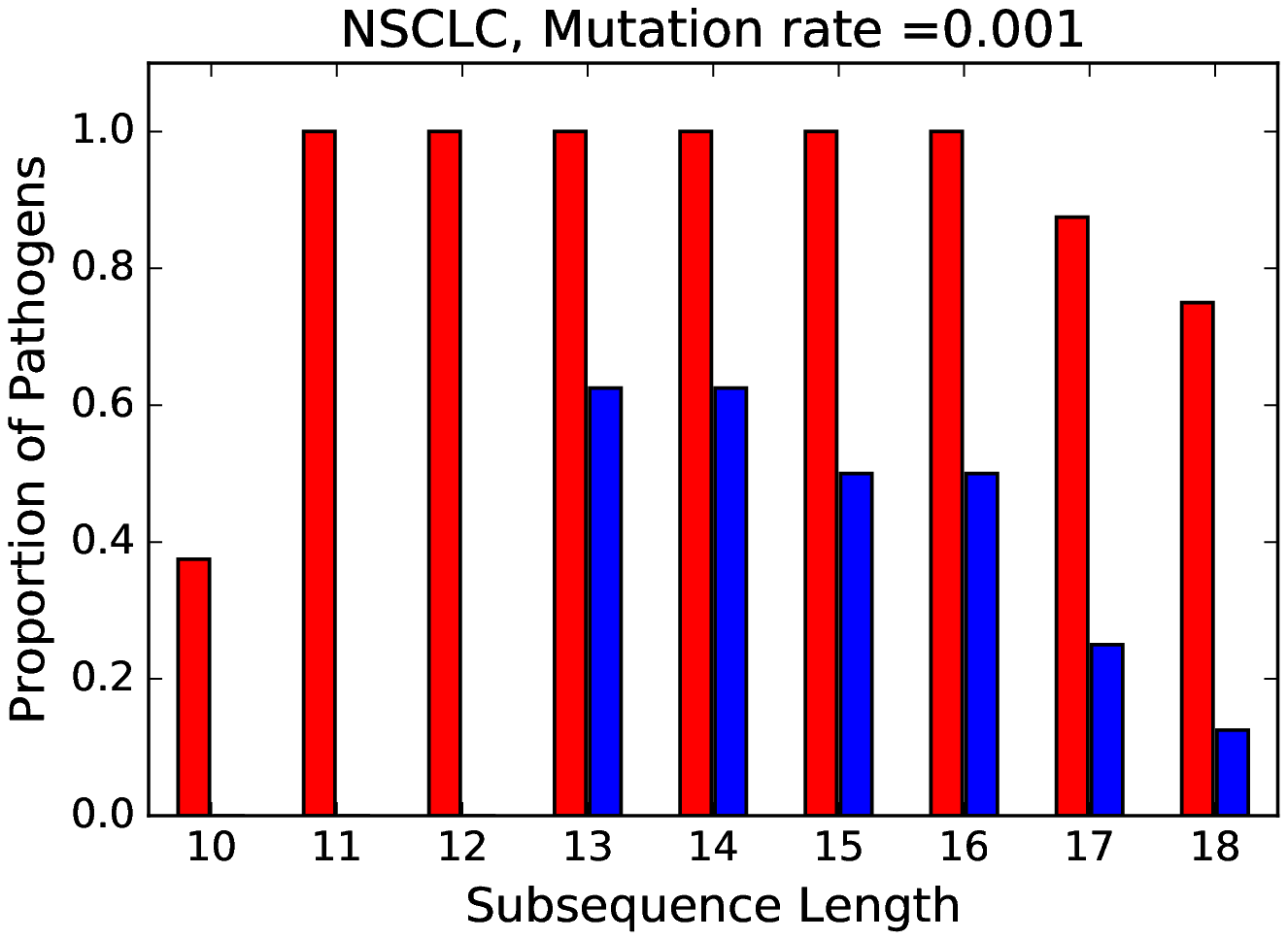}
       \includegraphics [width=0.32\textwidth]{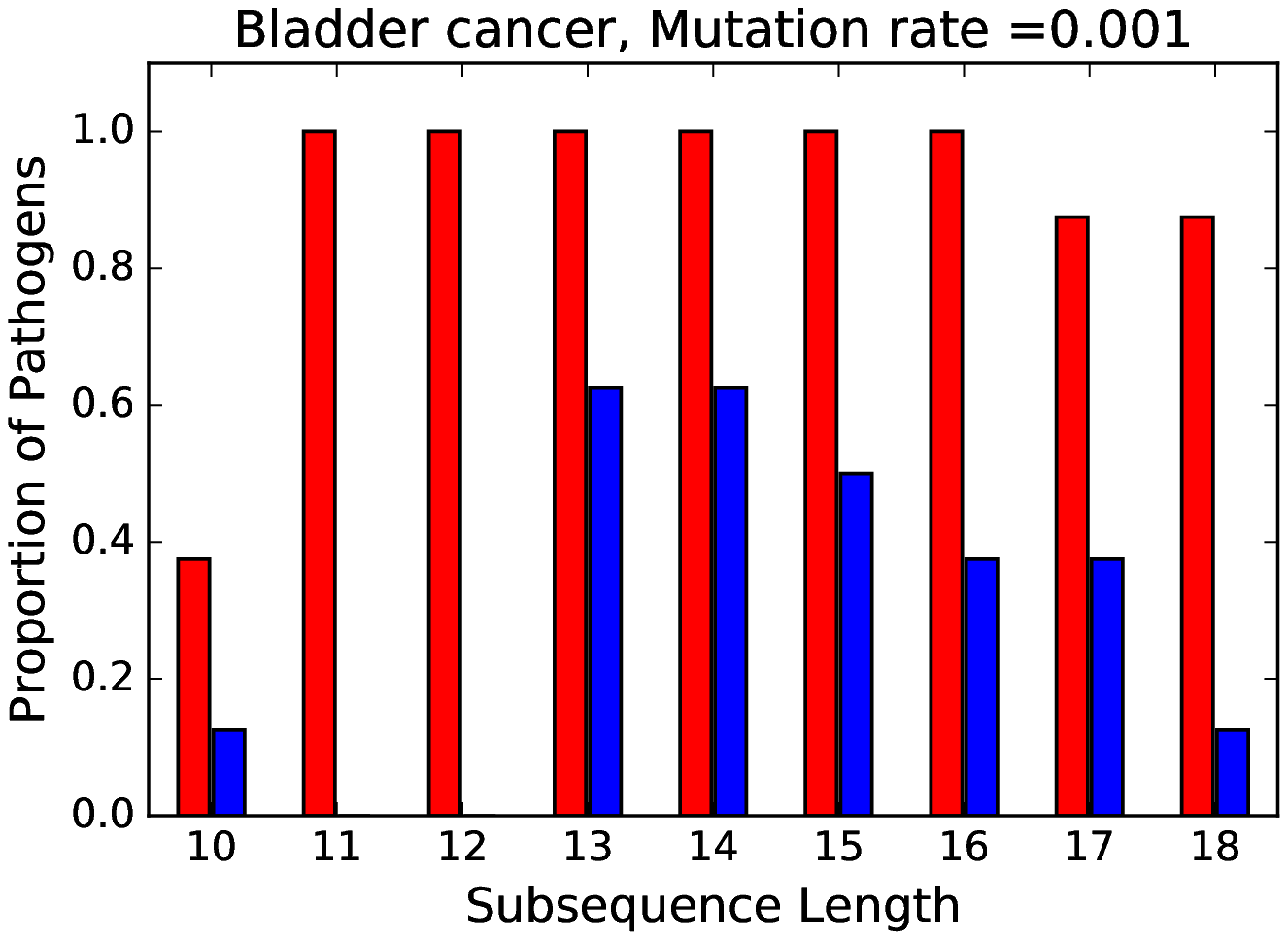}
            \includegraphics[width=0.32\textwidth]{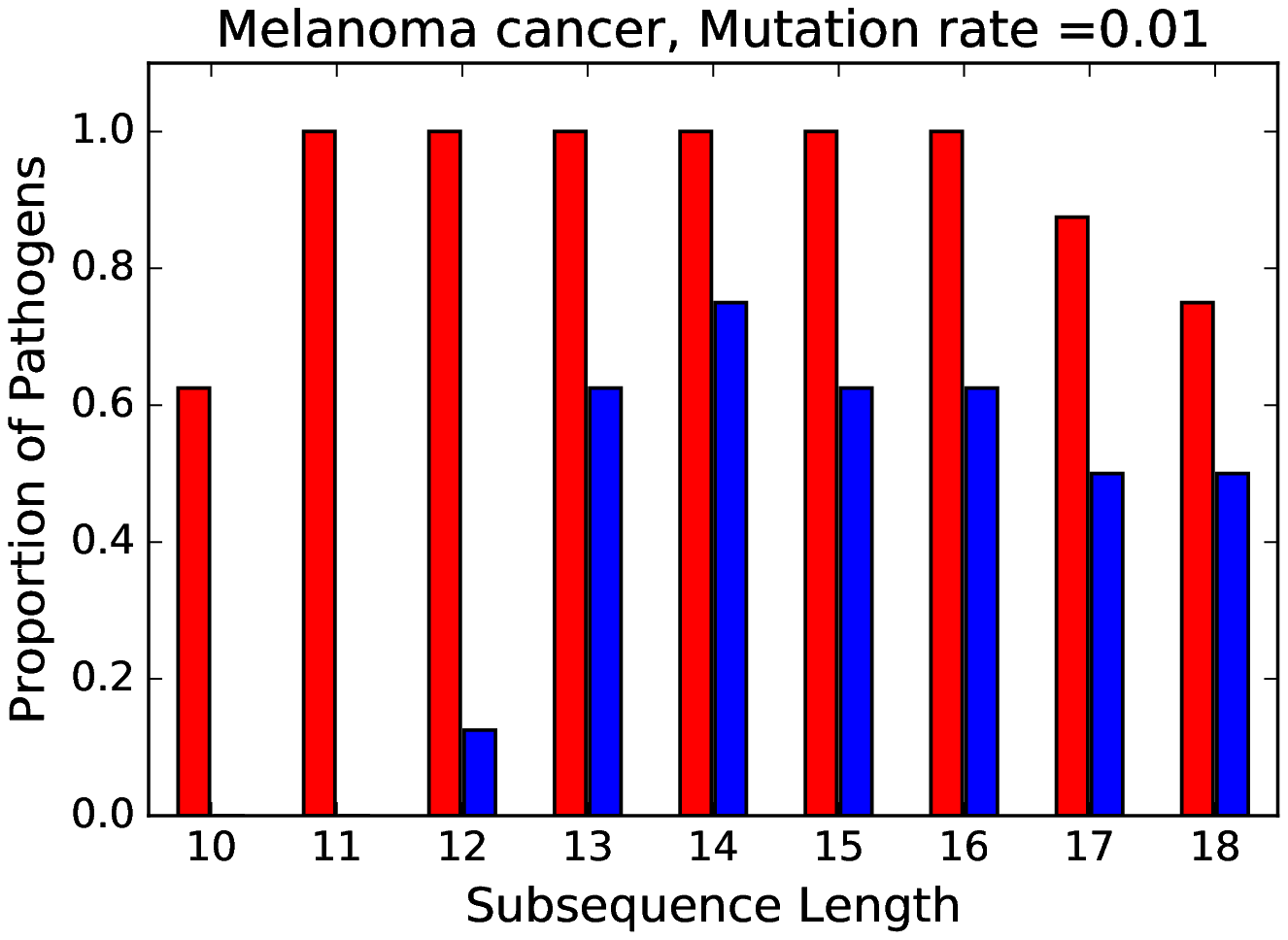}
    \includegraphics [width=0.32\textwidth]{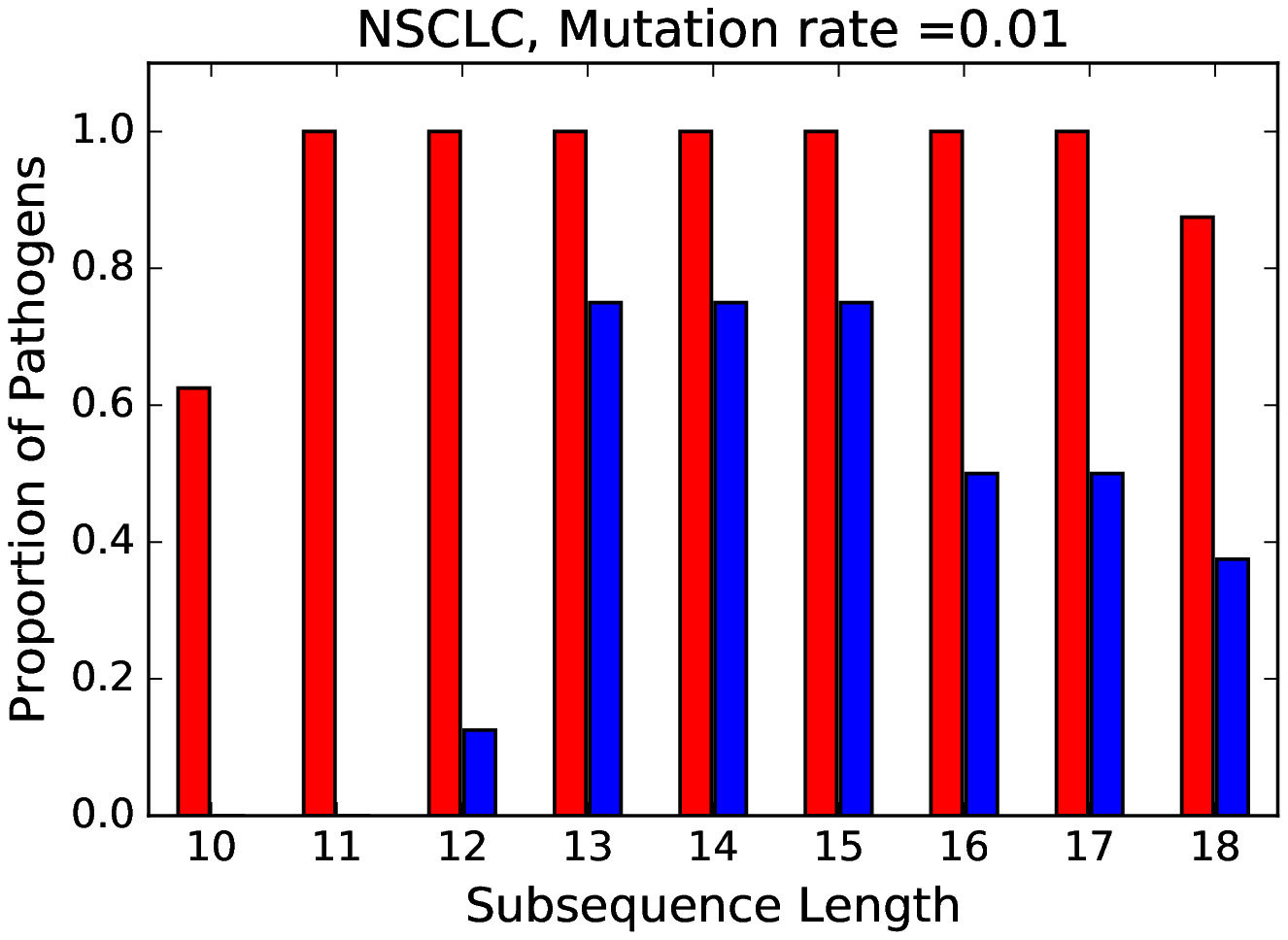}
      \includegraphics [width=0.32\textwidth]{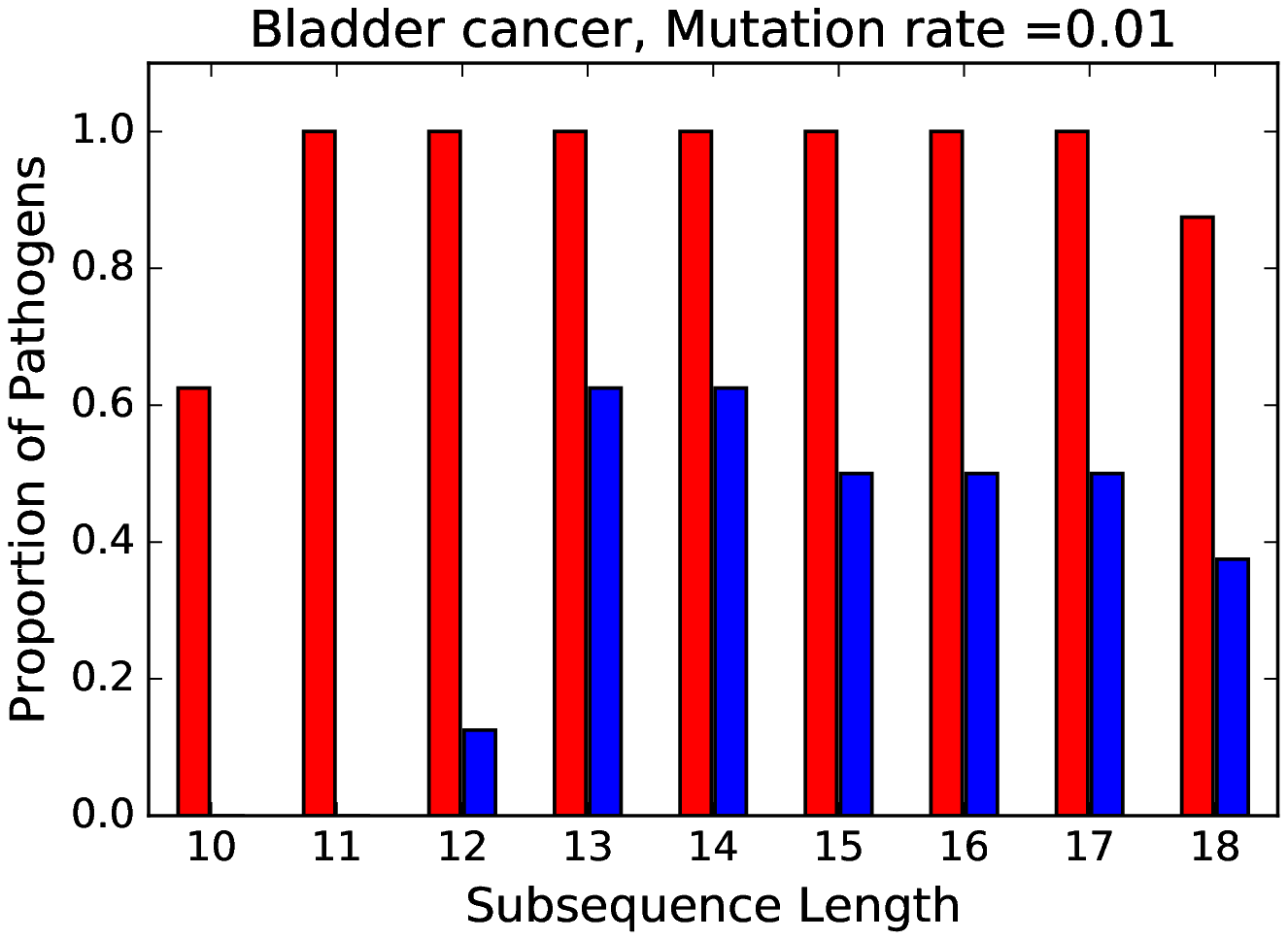}
        
\caption{ The height of the red bars represents the proportions of pathogen  DNA that are more similar to cancer exomes than to normal exome (one-sample t-test results with $p$-value $\leq 0.01$). The height of the blue bars represents the proportion of pathogens whose DNA are more similar to cancer exomes than to exomes with equal mutation rate but uniformly distributed mutations (two-sample one-sided t-test $p$-value $\leq 0.01$).}
        \label{fig:plot3}
\end{figure}

In Fig.~\ref{fig:plot3}, each histogram refers to a particular cancer and mutation rate. Red bars refer to Test $1$ and blue bars refer to Test $2$. Bar height represents, for any given subsequence length $\ell\in \{10,11,\ldots,18\}$, the  proportion of pathogens (out of the $8$ considered in this paper) for which the $p$-value is $\leq 0.01$. Related data can be found in the tables of the Appendices \ref{r1} \ref{r2}, and \ref{r3} for $\rho=0.0005$, $\rho=0.001$, and $\rho=0.01$, respectively. In these tables, the second column refers to $s_\ell(x,{y})$, the third column gives a $95\%$ confidence interval for $s_\ell(x,\tilde{y})$, the fourth  column gives the $p$-value for Test $1$ and the fifth column gives the $p$-value for Test $2$.
We make the following observations:
\begin{itemize}
\item[$\diamond$]
Referring to Test $1$ (red bars in Figs.~\ref{fig:plot3}), all three mutagenic processes render the human exome more similar to all pathogen DNA sequences at all $\rho\in \{0.0005, 0.001, 0.01\}$ and $\ell \in \{12,\ldots,16\}$. For $\ell\leq 11$ or $\ell\geq17$ the effect of the mutagenic processes on the similarity scores are less conclusive. This suggests that the increase of similarity is particularly relevant in the range of peptide sizes ($4-5$ amino-acids) that are relevant for epitope presentation in the human MHC presentation. Note, however, that the changes in similarity are small, typically $\ll 1\%$ (see tables in Sections ~\ref{r1}-\ref{r3}, Columns $2,3$).  

 \item[$\diamond$] Whether the above change of similarity is due to the specificity of the mutation distribution or random mutations trigger the same effect depends on the pathogen, the length, and the mutation rate. For instance, for Melanoma at $\ell=13$ the change in similarity due to cancer specific mutations is more pronounced for $5$ out of the $8$ pathogens, for $\rho\in \{0.0005, 0.001,  0.01\}$. By contrast, for all mutagenic processes there appears to be no statistical difference at length $11$. 
\end{itemize}
\subsubsection*{Impact of mutation rate}

\begin{figure}
\centerline{
\includegraphics[scale=0.31]{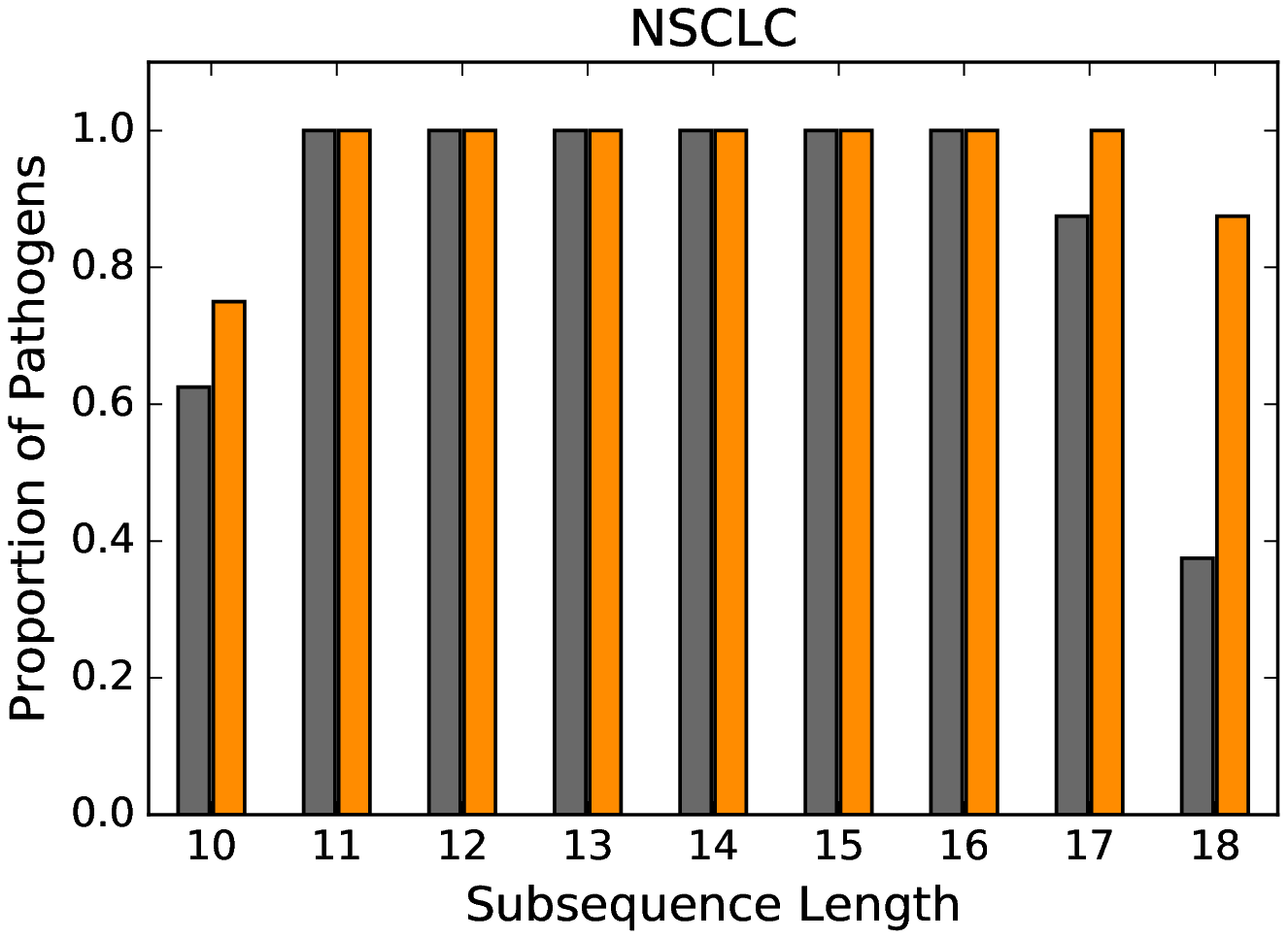} 
 \includegraphics[scale=0.31]{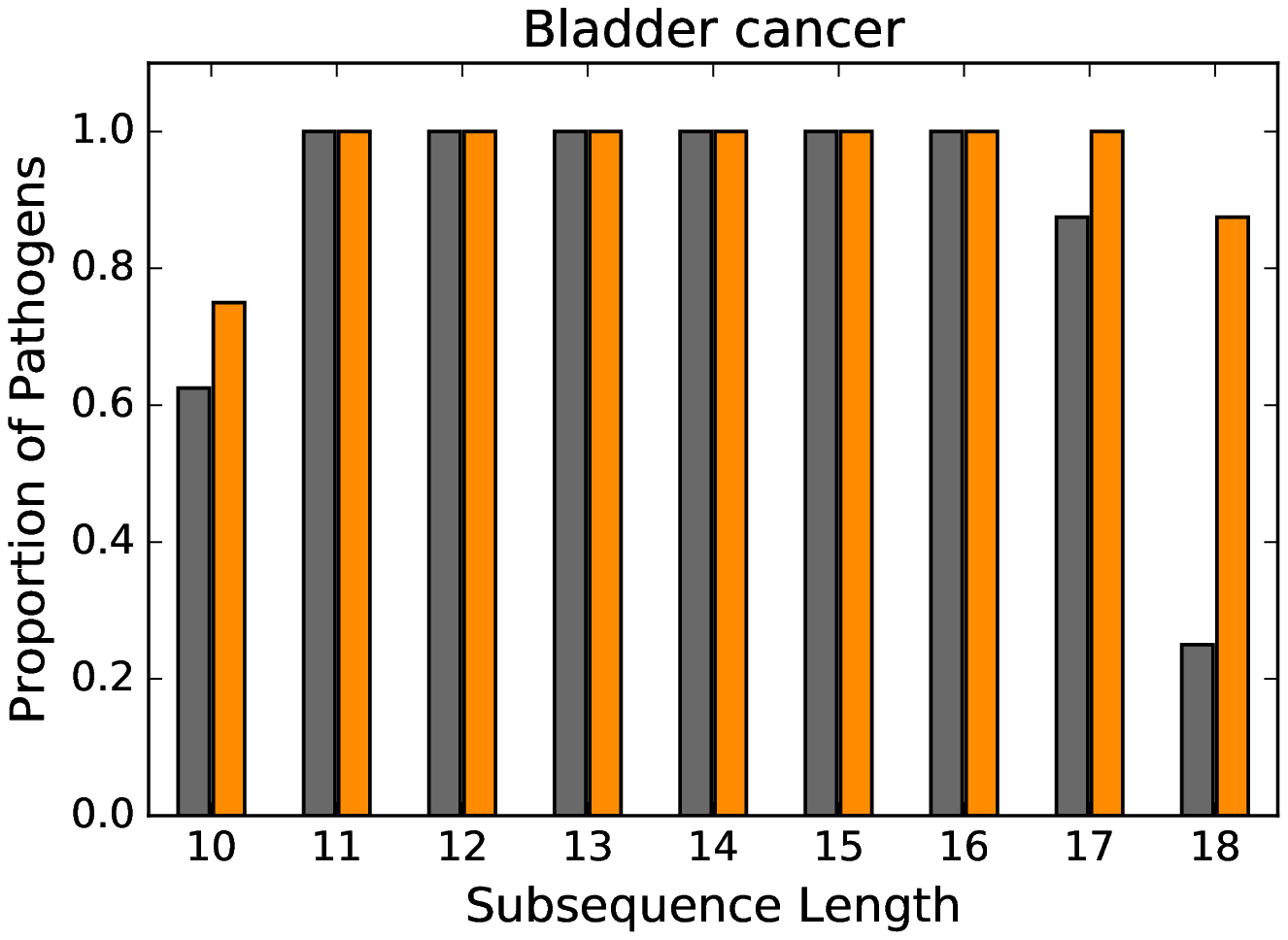}  
\includegraphics[scale=0.31]{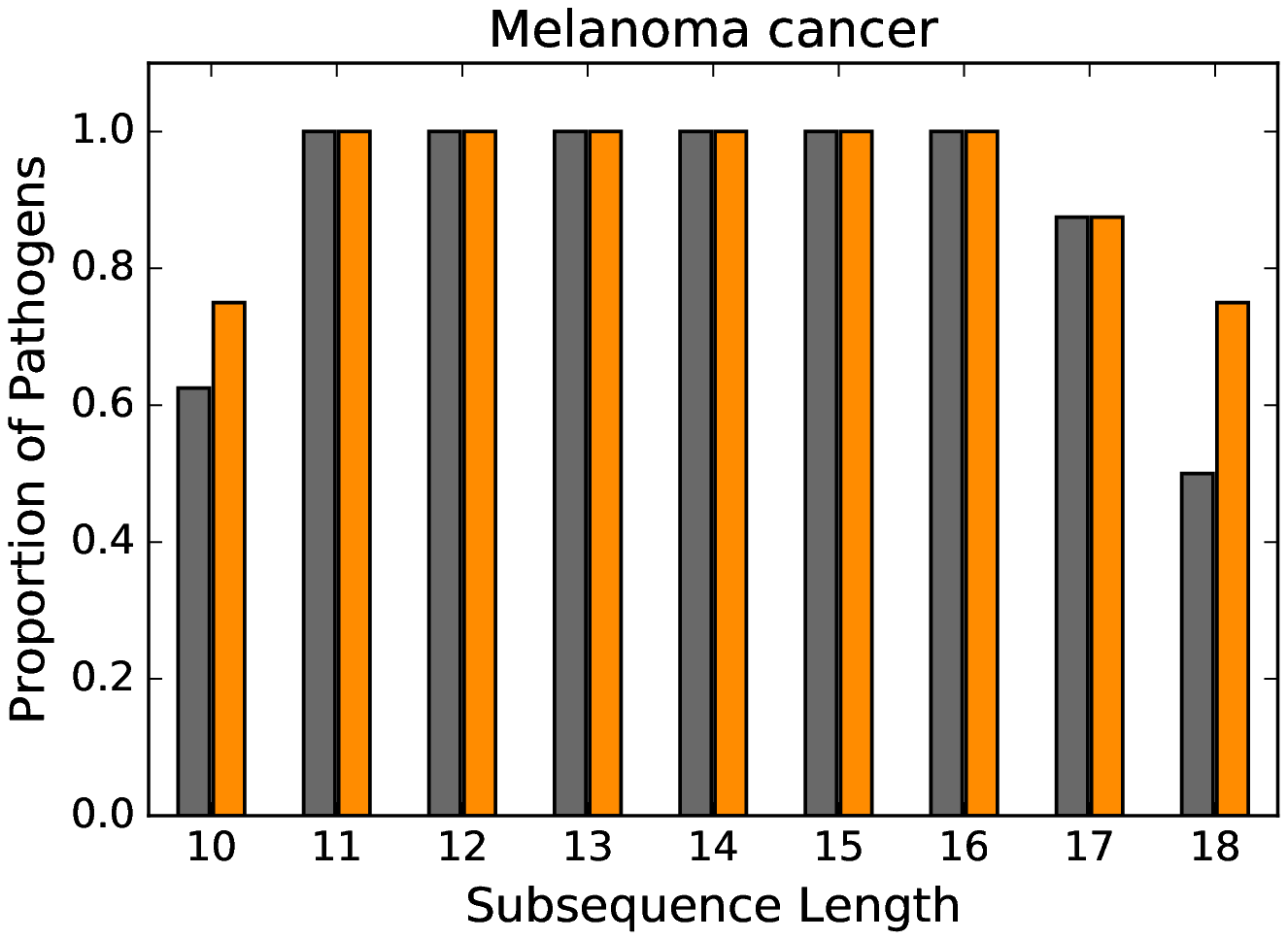}
 }
 \caption{The height of the grey bars represents the  proportion of pathogen DNA for which increasing mutation rate from $0.0005$ to $0.001$ results in an increased similarity with the cancer exome ($p$-value$\leq 0.01$). Orange bars refer to the same proportions but when mutation rate increases from $0.001$ to $0.01$. }
    \label{fig:plot4}
\end{figure}

To assess the impact of mutational rate on the similarity between pathogen DNA and human exome, for any given mutagenic process, pathogen DNA, and length we proceeded as follows.  We first generated $1000$ cancer exomes at mutation rate $\rho=0.0005$ and $1000$ cancer exomes at mutation rate $0.001$. Second, we computed the similarity scores of the two sets of cancer exomes relative to the pathogen DNA.   $P$-values were computed for comparing the means of the two sets of similarity scores using a two-sample one-sided t-test with a null hypothesis that the true mean of the similarity scores at the lowest rate ($\rho=0.0005$) is no larger than the true mean of the similarity scores at the higher rate ($\rho=0.001$).
We then repeated the experiment for  $\rho=0.001$ vs. $\rho=0.01$. In Fig.~\ref{fig:plot4}, the histograms represent the proportion of pathogens for which the $p$-value is $\leq 0.01$---grey bars refers to the $0.0005$ v.s. $0.001$ experiment and the orange bars refer to the $0.001$ v.s. $ 0.01$ experiment. We obtain the following result:
\begin{itemize}
\item[$\diamond$]
For all combinations of mutagenic processes and pathogens, and for all $\ell\in \{11,\ldots,16\}$, a higher mutation rate results in higher similarity score. For $\ell\in \{9,10,17,18\}$ results are inconclusive.
\end{itemize} 

\subsection{Resiliency of exomes with respect to mutagenic processes}
\label{meth3}
\begin{figure}
\begin{center}
\input{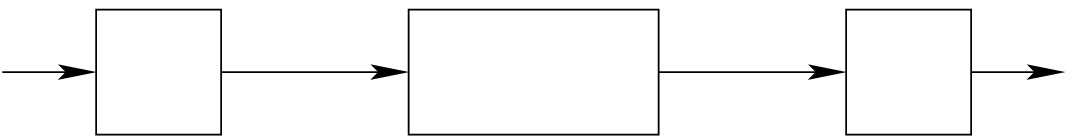}
\caption{\label{cancercha} Exome $\{y_1,y_2,\ldots\}$, which gives amino acid sequence $\{a_1,a_2,\ldots\}$,  undergoes specific somatic mutations through transformation ${\mathbb{P}}_{c,\rho}(\cdot|\cdot)$ and results in $\{\tilde{y}_1,\tilde{y}_2,\ldots\}$ which, in turn, gives amino acid sequence $\{\tilde{a}_1,\tilde{a}_2,\ldots\}$.}
\end{center}
\end{figure}
In order to compare the resiliency of the model organism exomes with respect to mutagenic processes, we evaluated the error correction capabilities of the genetic code (the codon allocation to amino-acids) for each combination of model exome and mutagenic process. 
 \begin{figure}
\centerline{
 \includegraphics[scale=0.35]{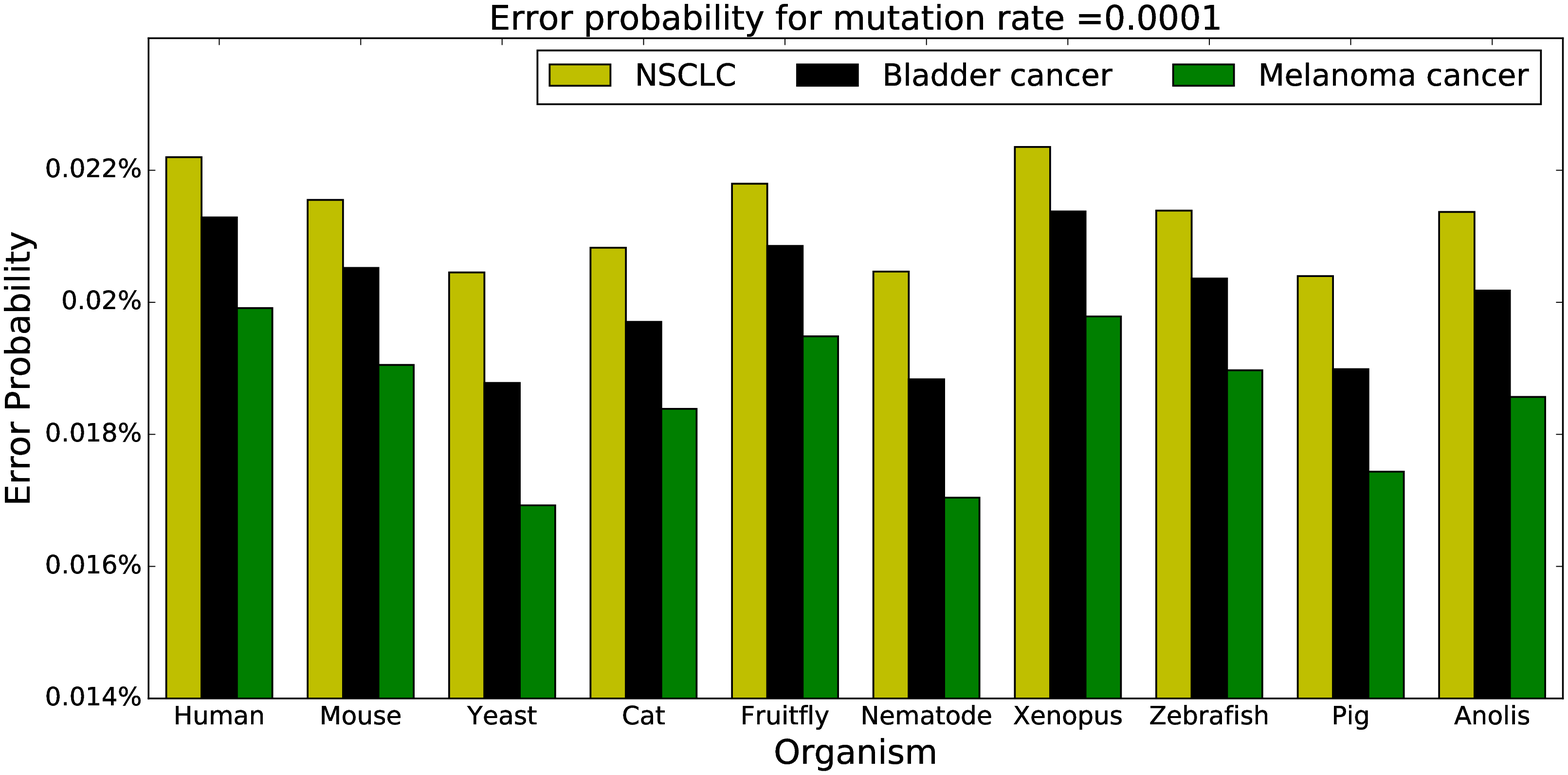}
 }
\centerline{
    \includegraphics[scale=0.35]{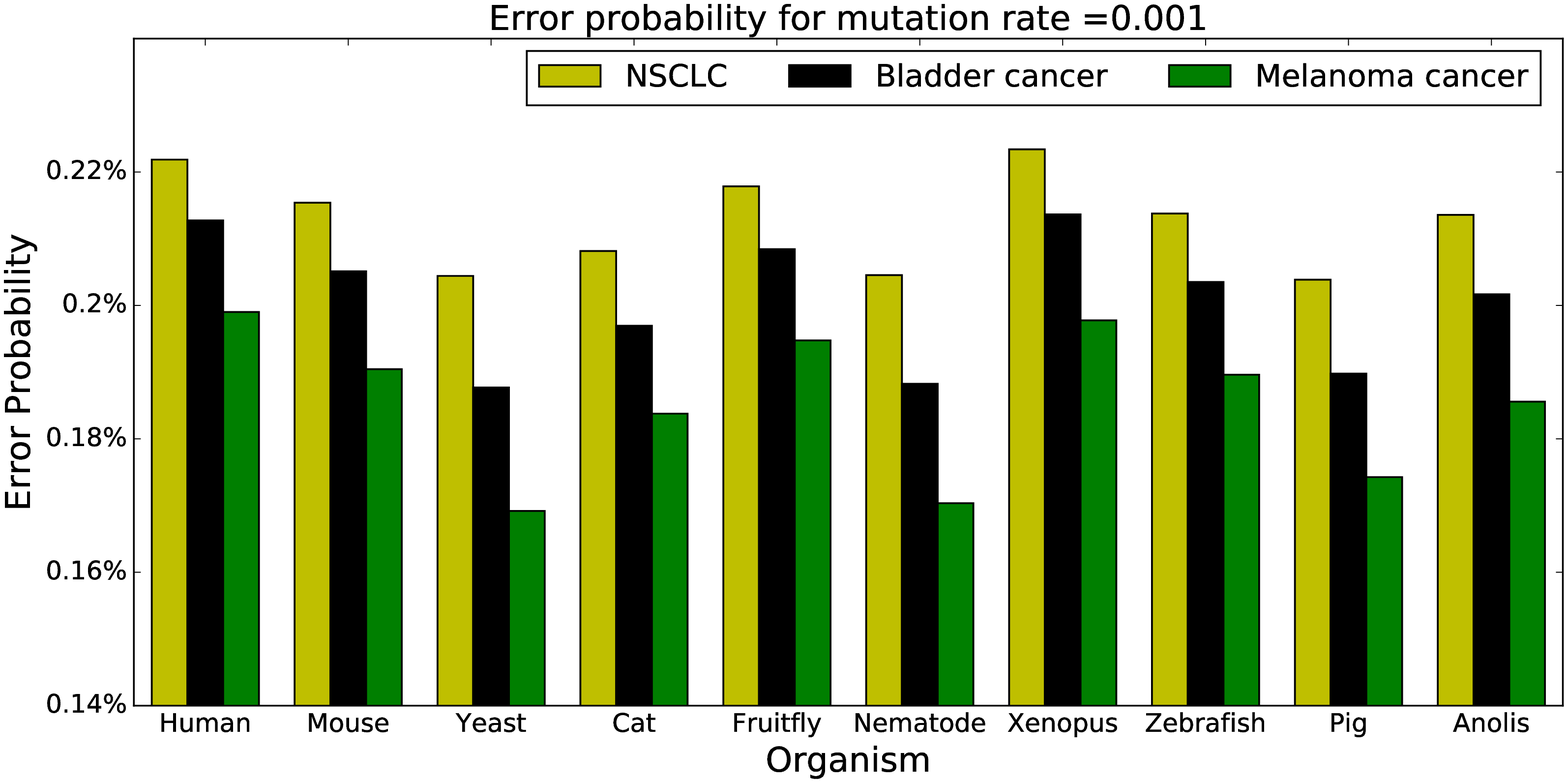}
   }
\centerline{    \includegraphics[scale=0.35]{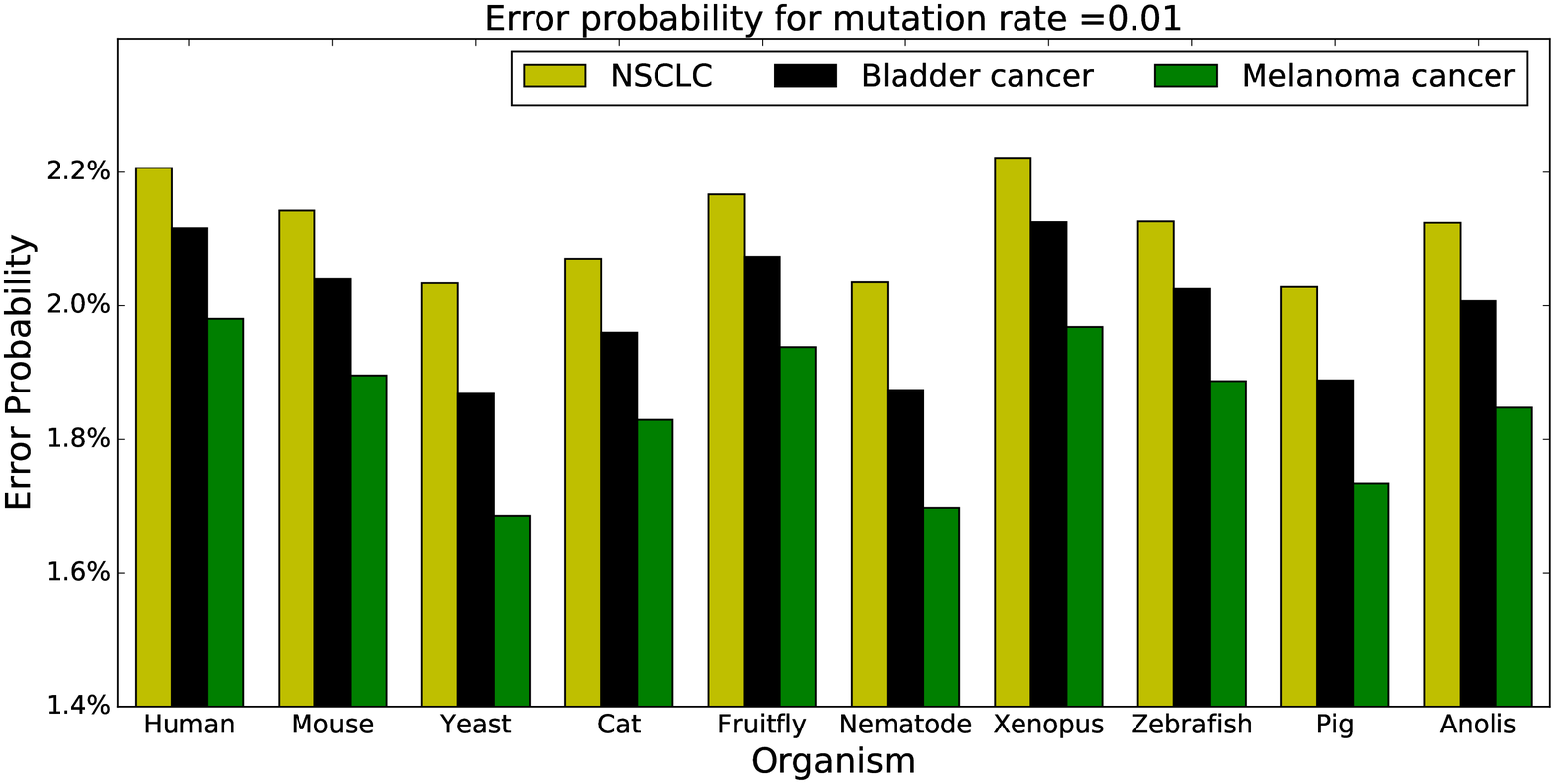}}
 \caption{Average proportions of erroneous amino acids after passing exomes through different cancer channels at mutation rater $0.0001$, $0.001$, and $0.01$.}
    \label{fig:plot2}
\end{figure}
Referring to Fig.~\ref{cancercha}, $y=\{y_1,\ldots,y_L\}$ represents a DNA sequence whose corresponding sequence of amino acids is $\{a_1,\ldots,a_{L/3}\}$. This DNA sequence is then passed through a given cancer channel $\mathbb{P}_{c,\rho}(\cdot|\cdot)$ and results in a cancer sequence $\tilde{y}=\{\tilde{y}_1,\tilde{y}_2,\ldots,\tilde{y}_L\}$ and a corresponding sequence of cancer amino acids $\{\tilde{a}_1,\tilde{a}_2,\ldots,\tilde{a}_{L/3}\}$. From $\{a_1,a_2,\ldots,{a}_{L/3}\}$ and $\{\tilde{a}_1,\tilde{a}_2,\ldots,\tilde{a}_{L/3}\}$  we computed the relative proportion of amino acids that were affected, that is  \begin{align}\frac{|\{i: \tilde{a}_i\ne a_i\}|}{(L/3) }.\label{prerror}
\end{align}
Finally, averaging over all possible realizations of $\tilde{y}$ (and therefore over $\tilde{a}$), we obtained the average error probability 
\begin{align}\label{per}
\mathbb{P}(\text{error}| y,c,\rho)=\frac{{\mathbb{E}}|\{i: \tilde{a}_i\ne a_i\}|}{(L/3) }.
\end{align}
Fig.~\ref{fig:plot2} represents $\mathbb{P}(\text{error}| y,c,\rho)$ for each combination of model organism, cancer mutation process, and mutation rate $\rho\in \{0.0001, 0.001, 0.01\}$. Notice that $\mathbb{P}(\text{error}| y,c,\rho)$ is not a linear function of $\rho$. 
Computation details for $\mathbb{P}(\text{error}| y,c,\rho)$ are deferred to the Appendix~\ref{cb}. 
 Referring to Fig.~\ref{fig:plot2}, we obtain the following result:
\begin{itemize}
\item[$\diamond$]
Although the proportion of non-synonymous mutations varies across exomes for the three types of mutagenic processes, it is always lowest for melanoma and maximal for lung. Moreover, this ordering holds irrespectively of the intensity of the mutation rate. It should be noted that we evaluated the proportions of non-synonymous mutations for several other organisms as well (including the set of pathogens considered in this paper) and this finding was validated in all cases.  
\end{itemize}
\section{Discussion}

We employed large scale simulations to model the random (across space) effect of stochastic mutagenic processes on the human normal genome.  We believe this is a valid approach since the cancer exome available data does suggest that, while at the granular level mutation rates vary, the mutagenic processes in cancers with large number of mutations affect equally all chromosomal regions of the exome \cite{lawrence2013mutational}.  Essentially, we simplify the analysis using this assumption.

Our in-silico results show that, in general, the typical stochastic mutagenic processes encountered in the major cancer indications with abundant neoantigens do appear to shift the peptide distribution of the modified exome universally towards a landscape that appears more similar to pathogenic insult. Specifically, all three mutagenic processes considered induce subtle but robust shifts in the measure by which we characterized the similarity between the normal human exome and pathogen DNA sequences, at mutation rates in the upper range of the mutation rates observed in actual cancer samples ($\geq 0.0005$). Moreover, the range of peptide lengths where this shift happens aligns with the typical length of  peptides presented by the human MHC presentation system, suggesting an increased potential for recognition of these types of somatic mutations by a pathogen-trained host immune system.

We also note that for many combinations of pathogen DNA and mutagenic process cases this increase of similarity cannot be solely attributed to the mutation distribution; randomly and uniformly distributed mutations can cause similar shifts in similarity. By contrast, increasing the mutation rate while keeping the underlying mutation distribution fixed always results in an increased similarity betweeen human exome and pathogen DNA at $\ell\in \{11,\ldots,16\}$, which again corresponds to the length of peptides presented by the human presentation system. This suggests that the intensity of the mutational rate is an important parameter that directly affects the similarity between cancer exome and pathogen DNA.

We also observe that the effect of the considered mutagenic processes on the likelihood of observing a non-synonymous alteration is strikingly different across processes but consistent across the species studied in our framework (human and model organisms). Melanoma/UV light  alterations are the least likely to result in amino acid functional changes, followed by APOBEC-driven alterations and then by smoking alterations, suggesting different error-correcting capabilities of the living exomes towards this various mutagenic insults.  This is an attractive observation from an evolutionary perspective: due to universal exposure to sunlight, organisms likely developed similarly universal intrinsic protection from UV light type of modifications to their exomes via the redundancies in the aminoacid codon allocation.  Similarly, APOBEC-activation appears to be a universal innate protection mechanism that allows the cell to induce damaging mutations to foreign organisms, while the mutations resulting from tobacco smoking are less likely to have presented evolutionary pressure. 
In summary, our in-silico approach reveals two competing mechanisms of tolerance pressure on the major mutagenic processes present in human cancers that modulate the potential immune recognition of alterations at the exome level through pathogen similarity and through functional redundancy; the balance between these mechanisms may significantly contribute to the eventual mutational landscape of advanced cancers.

\appendix
\section{Appendices}\label{support}

\subsection{Data for Fig.~\ref{fig:plot1}} \label{app1}

\label{supportaa}

In the table below we listed the similarity scores $s_\ell(x,y)$ of each pathogen $x$ against the human exome $y$, as a function of the subsequence length $\ell$.
\subsubsection{Matching score of pathogens against human exome}
\label{matchoscore}
\vspace{.5cm}
\centerline{\begin{tabular}{ |c | c | c | c | c | c | }
\hline 
  $\ell$   &Ebola virus	&CMV	&Dengue virus	&EBV	&HHV	\\
 \hline 
9	&100.0	&100.0	&100.0	&100.0	&100.0	\\
10	&99.94	&99.71	&100.0	&99.94	&99.83	\\
11	&98.11	&94.88	&99.30	&98.26	&95.96	\\
12	&86.99	&74.92	&91.65	&86.74	&79.24	\\
13	&56.32	&43.52	&64.19	&58.29	&49.20	\\
14	&23.94	&18.64	&29.87	&27.43	&21.84	\\
15	&7.82	&6.60	&9.98	&10.05	&7.91	\\
16	&2.40	&2.23	&2.81	&3.19	&2.73	\\
17	&0.62	&0.77	&0.73	&1.02	&1.05	\\
18	&0.12	&0.27	&0.19	&0.33	&0.48	\\
 \hline 
   \end{tabular}}
\vspace{.5cm}
\centerline{\begin{tabular}{ |c | c | c | c | c | c | }
\hline 
  $\ell$   &HPV	&Measles virus	&Yellow fever virus	&Random	\\
 \hline 
9	&100.0	&100.0	&100.0	&100$\pm0.002$\\
10	&99.97	&99.97	&100.0	&100$\pm0.002$\\
11	&99.05	&98.23	&99.05	&96.4$\pm0.002$\\
12	&90.48	&86.15	&92.10	&79.4$\pm0.002$\\
13	&61.24	&54.27	&65.28	&48.3$\pm0.002$\\
14	&28.22	&22.64	&30.54	&20$\pm0.002$\\
15	&9.84	&7.29	&10.81	&6.2$\pm0.002$\\
16	&3.19	&2.19	&3.15	&1.116$\pm0.001$\\
17	&1.10	&0.68	&0.87	&0.28$\pm0.001$\\
18	&0.50	&0.19	&0.29	&0.07$\pm0.001$\\
 \hline 
   \end{tabular}}
\vspace{.5cm}
The column ``Random'' refers to a $95\%$ confidence interval for the similarity score between  a randomly generated pathogen sequence $X$, where each nucleotide is independently and uniformly selected with probability $1/4$, and the normal human exome $y$. To compute this confidence interval we proceeded as follows. The similarity score for a random instance $X$ of length $L$ is given by
$$s_\ell(X,y)= \frac{1}{L-\ell +1}\sum_{i=1}^{L-\ell+1}Z_i$$
where the $Z_i$'s are i.i.d. Bernoulli random variables such that 
\begin{align}
\label{bernp}Pr(Z_i=1)=1-Pr(Z_i=0)=\min\{1,M_\ell/4^\ell\}\defeq s_\ell(y).\end{align} 
Here $M_\ell$ denotes the number of distinct length-$\ell$ substrings in the human genome and was
computed empirically for $\ell\in \{9,10,\ldots,15\}$:
$$M_9 = \SI{262144}{} \quad M_{10} = \SI{1047125} \quad M_{11} = \SI{4043826}\,$$ $$M_{12} = \SI{13319884}\quad M_{13} = \SI{32427184} \quad M_{14} = \SI{53660993} \quad M_{15} = \SI{66828678}\,$$ $$\quad M_{16} = \SI{47945739}\quad M_{17} = \SI{49134120}\quad M_{18} = \SI{49785872}.$$

Taking expectation over $X$ yields
$$\mathbb{E} (s_\ell(X,y))=EZ_i= s_\ell(y).$$

A confidence inteval for $s_\ell(X,y)$ was computed via Chebyshev's inequality as follows. We have
\begin{align}\label{cheb}Pr(|s_\ell(X,y)-s_\ell(y) |\geq \varepsilon_\ell)\leq \frac{\text{Var} (s_\ell(X,y))}{\varepsilon_\ell^2}=\frac{\text{Var}(\sum_{i=1}^{L-\ell+1}Z_i)}{(L-\ell+1)^2\varepsilon_\ell^2}.
\end{align}
Furthermore,
\begin{align*}
\text{Var}(\sum_{i=1}^{L-\ell+1}Z_i)&=\sum_{i=1}^{L-\ell+1}\text{Var}(Z_i)+2 \sum_{i<j}\text{Cov}(Z_i,Z_j)\nonumber \\
&=(L-\ell+1)\text{Var}(Z_1)+2\sum_{i=1}^{L-\ell+1}\sum_{j=i+1}^{i+\ell-1}\text{Cov}(Z_i,Z_j)
\end{align*}
where for the second equality we used the fact that the $Z_i$'s are identically distributed and that $Z_k$ and $Z_j$ are independent whenever $j\geq k+\ell$. 
Now $$\text{Var}(Z_1)=s_\ell(y)(1-s_\ell(y))\leq s_\ell(y)$$ and since the $Z_i$'s are binary random variables
\begin{align*}
\text{Cov}(Z_i,Z_j)&=E(Z_iZ_j)-E(Z_i)E(Z_j)\notag \\
&\leq E(Z_i)-E(Z_i)E(Z_j)\notag \\
&=s_\ell(y)(1-s_\ell(y))\nonumber \\
&\leq s_\ell(y).
\end{align*}
Therefore,
\begin{align}\label{lavar}
\text{Var}(\sum_{i=1}^{L-\ell+1}Z_i)\leq s_\ell(y)(L-\ell+1)(1+2(\ell-1)).
\end{align}
Finally, from \eqref{bernp}, \eqref{cheb}, and \eqref{lavar} we get
$$  Pr(|s_\ell(X,y)-s_\ell(y)|\geq \varepsilon_\ell)\leq \frac{s_\ell(y)(1+2(\ell-1))}{(L-\ell+1)\varepsilon_\ell^2}=  \frac{\min\{1, M/4^\ell\}(1+2(\ell-1))}{(L-\ell+1)\varepsilon_\ell^2}. $$
To obtain a $95\%$ confidence interval we picked 
\begin{align}
\varepsilon_\ell=\left(\frac{\min\{1, M/4^\ell\}(1+2(\ell-1))}{(L-\ell+1)0.05}\right)^{1/2}
\end{align}
which is below $ 0.002$ for all $\ell \in \{9,10,\ldots,15\}$ regardless of the pathogen length $L$.

\subsection{Cancer channel}\label{appB}
We describe how we obtained cancer channel ${\mathbb{P}}_{c,\rho}(\cdot|\cdot)$ for a given cancer and mutation rate.
  For each cancer $c$ (Melanoma cancer, NSCLC, Bladder cancer)
we considered the set ${\cal{S}}_c$ of patients in \cite[Supplementary information, Table S2]{lawrence2013mutational} with that cancer. Then, for every mutation $\alpha\to \beta$ we empirically computed the average proportion of mutations across patients $$p_c({\alpha\to \beta})\defeq \frac{1}{|{\cal{S}}_c|}\sum_{i\in {\cal{S}}_c} p_c({i,\alpha\to \beta})$$ where $p_c({i,\alpha\to \beta})$ denotes the proportion of $\alpha\to \beta$ mutations among all mutations in patient $i$ and was computed from \cite[Supplementary information, Table S2]{lawrence2013mutational}. The probability that a nucleotide $\alpha$ in the normal exome results in nucleotide $\beta$ in the cancer exome is therefore given by  
$${\mathbb{P}}_{c,\rho}(\beta|\alpha)=\frac{p_c({\alpha\to \beta})\rho}{p(\alpha)}$$
for $\beta\ne \alpha$ and
$${\mathbb{P}}_{c,\rho}(\alpha|\alpha)=1-\frac{\rho}{p(\alpha)}\sum_{\beta\ne \alpha}p_c({\alpha\to \beta}).$$
The parameter $\rho$ denotes the overall mutation rate and $p(\alpha)$ denotes the relative number of nucleotide $\alpha$ in the exome and was computed from \cite[Supplementary information, Table S2]{lawrence2013mutational}.

\begin{rem}Because in the data from \cite[Supplementary information, Table S2]{lawrence2013mutational} complementary mutations were counted under the same category ({\it{e.g.}}, a change from cytosine to tyamine would be treated the same as a change from guanine to adenine), mutation types were considered in pairs. Since the relative proportions of complementary pairs were not given inf, we made the assumption that they were equal. Hence, in the above expression $p_c(i,\alpha \to \beta, i)$ actually corresponds to $$p_c(i,\alpha \to \beta, i)/2+p_c(i,\alpha' \to \beta', i)/2$$ where $(\alpha',\beta')$ is the complementary pair of $(\alpha,\beta)$.
\end{rem}

The second column in the tables of Sections~\ref{r1}-\ref{r3} represents $s_\ell(x,y)$ as a function of $\ell$. The third column represents a $95\%$ confidence interval for $s_\ell(x,\tilde{y})$ obtained through a standard application of the central limit theorem. This confidence interval is given by  
$$\Big[s_\ell(x)\pm \frac{1.96 \sigma}{\sqrt{1000}}\Big],$$ where $s_\ell(x)$ denotes the average of $s_\ell(x,\tilde{y})$ over the $1000$ independent trials $\{\tilde{y}_i\}_{i= 1}^{1000}$ and where $\sigma$ denotes the empirical standard deviation of $s_\ell(x,\tilde{y})$. The fourth column in the tables of Sections~\ref{r1}-\ref{r3} gives the $p$-value for Test $1$ and the fifth column gives the $p$-value for Test $2$.

\subsection{$\rho=0.0005$}\label{r1}

\subsubsection{Ebola virus indication}

\centerline{
\begin{tabular}{ |c | c | c | c | c |}
 \hline
  $\ell$  & Normal & Lung, $\rho=0.0005$& P-value& P-value random \\
    \hline
9	&100.0	&99.99±$(<10^{-5})$	&1.0	&0.5  \\
10	&99.94	&99.94±0.0002	&1.0	&$<0.01$  \\
11	&98.11	&98.13±0.0011	&$<0.01$	&0.9999  \\
12	&86.99	&87.08±0.0022	&$<0.01$	&0.8762  \\
13	&56.32	&56.45±0.0032	&$<0.01$	&$<0.01$  \\
14	&23.94	&24.03±0.0027	&$<0.01$	&$<0.01$  \\
15	&7.82	&7.86±0.0018	&$<0.01$	&$<0.01$  \\
16	&2.40	&2.41±0.0010	&$<0.01$	&$<0.01$  \\
17	&0.62	&0.63±0.0005	&$<0.01$	&$<0.01$  \\
18	&0.12	&0.12±0.0002	&$<0.01$	&0.4808  \\
\hline
 \end{tabular}}

\vspace{.5cm}

\centerline{
\begin{tabular}{ |c | c | c | c | c |}
 \hline
  $\ell$  & Normal & Bladder, $\rho=0.0005$& P-value& P-value random \\
    \hline
9	&100.0	&99.99±$(<10^{-5})$	&1.0	&0.5  \\
10	&99.94	&99.94±0.0002	&1.0	&$<0.01$  \\
11	&98.11	&98.14±0.0012	&$<0.01$	&0.9872  \\
12	&86.99	&87.08±0.0022	&$<0.01$	&0.8483  \\
13	&56.32	&56.45±0.0031	&$<0.01$	&$<0.01$  \\
14	&23.94	&24.02±0.0027	&$<0.01$	&$<0.01$  \\
15	&7.82	&7.85±0.0017	&$<0.01$	&$<0.01$  \\
16	&2.40	&2.41±0.0010	&$<0.01$	&$<0.01$  \\
17	&0.62	&0.63±0.0005	&$<0.01$	&0.0831  \\
18	&0.12	&0.12±0.0003	&$<0.01$	&0.5661  \\
\hline
 \end{tabular}}

\vspace{.5cm}

\centerline{
\begin{tabular}{ |c | c | c | c | c |}
 \hline
  $\ell$  & Normal & Mela, $\rho=0.0005$& P-value& P-value random \\
    \hline
9	&100.0	&99.99±$(<10^{-5})$	&1.0	&0.5  \\
10	&99.94	&99.94±0.0002	&1.0	&0.0706  \\
11	&98.11	&98.13±0.0010	&$<0.01$	&1.0  \\
12	&86.99	&87.07±0.0021	&$<0.01$	&0.9999  \\
13	&56.32	&56.46±0.0032	&$<0.01$	&$<0.01$  \\
14	&23.94	&24.03±0.0027	&$<0.01$	&$<0.01$  \\
15	&7.82	&7.86±0.0018	&$<0.01$	&$<0.01$  \\
16	&2.40	&2.41±0.0010	&$<0.01$	&$<0.01$  \\
17	&0.62	&0.63±0.0005	&$<0.01$	&$<0.01$  \\
18	&0.12	&0.12±0.0003	&$<0.01$	&0.5378  \\
\hline
 \end{tabular}}

\vspace{.5cm}

\subsubsection{CMV indication}

\centerline{
\begin{tabular}{ |c | c | c | c | c |}
 \hline
  $\ell$  & Normal & Lung, $\rho=0.0005$& P-value& P-value random \\
    \hline
9	&100.0	&99.99±$(<10^{-5})$	&1.0	&0.8501  \\
10	&99.71	&99.72±0.0001	&$<0.01$	&0.9985  \\
11	&94.88	&94.94±0.0005	&$<0.01$	&1.0  \\
12	&74.92	&75.05±0.0008	&$<0.01$	&1.0  \\
13	&43.52	&43.63±0.0008	&$<0.01$	&1.0  \\
14	&18.64	&18.70±0.0006	&$<0.01$	&1.0  \\
15	&6.60	&6.62±0.0004	&$<0.01$	&1.0  \\
16	&2.23	&2.23±0.0002	&$<0.01$	&0.9999  \\
17	&0.77	&0.77±0.0001	&$<0.01$	&0.9994  \\
18	&0.27	&0.27±0.0001	&$<0.01$	&0.9999  \\
\hline
 \end{tabular}}

\vspace{.5cm}

\centerline{
\begin{tabular}{ |c | c | c | c | c |}
 \hline
  $\ell$  & Normal & Bladder, $\rho=0.0005$& P-value& P-value random \\
    \hline
9	&100.0	&99.99±$(<10^{-5})$	&1.0	&0.7875  \\
10	&99.71	&99.72±0.0001	&$<0.01$	&0.0189  \\
11	&94.88	&94.95±0.0005	&$<0.01$	&0.9999  \\
12	&74.92	&75.07±0.0008	&$<0.01$	&1.0  \\
13	&43.52	&43.64±0.0009	&$<0.01$	&1.0  \\
14	&18.64	&18.70±0.0007	&$<0.01$	&0.9999  \\
15	&6.60	&6.62±0.0004	&$<0.01$	&0.9999  \\
16	&2.23	&2.24±0.0003	&$<0.01$	&0.9760  \\
17	&0.77	&0.77±0.0001	&$<0.01$	&0.8426  \\
18	&0.27	&0.27±0.0001	&$<0.01$	&0.4867  \\
\hline
 \end{tabular}}

\vspace{.5cm}

\centerline{
\begin{tabular}{ |c | c | c | c | c |}
 \hline
  $\ell$  & Normal & Mela, $\rho=0.0005$& P-value& P-value random \\
    \hline
9	&100.0	&99.99±$(<10^{-5})$	&1.0	&0.1131  \\
10	&99.71	&99.72±0.0001	&$<0.01$	&0.9999  \\
11	&94.88	&94.94±0.0005	&$<0.01$	&1.0  \\
12	&74.92	&75.04±0.0008	&$<0.01$	&1.0  \\
13	&43.52	&43.62±0.0008	&$<0.01$	&1.0  \\
14	&18.64	&18.69±0.0007	&$<0.01$	&1.0  \\
15	&6.60	&6.61±0.0004	&$<0.01$	&1.0  \\
16	&2.23	&2.23±0.0002	&$<0.01$	&1.0  \\
17	&0.77	&0.77±0.0001	&$<0.01$	&0.9999  \\
18	&0.27	&0.27±0.0001	&$<0.01$	&0.7213  \\
\hline
 \end{tabular}}

\vspace{.5cm}

\subsubsection{Dengue virus indication}

\centerline{
\begin{tabular}{ |c | c | c | c | c |}
 \hline
  $\ell$  & Normal & Lung, $\rho=0.0005$& P-value& P-value random \\
    \hline
9	&100.0	&99.99±$(<10^{-5})$	&1.0	&0.5  \\
10	&100.0	&99.98±$(<10^{-5})$	&1.0	&0.9714  \\
11	&99.30	&99.29±0.0010	&0.9943	&0.9994  \\
12	&91.65	&91.69±0.0026	&$<0.01$	&0.7855  \\
13	&64.19	&64.29±0.0041	&$<0.01$	&0.1778  \\
14	&29.87	&29.94±0.0042	&$<0.01$	&$<0.01$  \\
15	&9.98	&10.01±0.0028	&$<0.01$	&0.0106  \\
16	&2.81	&2.82±0.0016	&$<0.01$	&0.0929  \\
17	&0.73	&0.74±0.0008	&$<0.01$	&0.2427  \\
18	&0.19	&0.19±0.0004	&$<0.01$	&0.9153  \\
\hline
 \end{tabular}}

\vspace{.5cm}

\centerline{
\begin{tabular}{ |c | c | c | c | c |}
 \hline
  $\ell$  & Normal & Bladder, $\rho=0.0005$& P-value& P-value random \\
    \hline
9	&100.0	&99.99±$(<10^{-5})$	&1.0	&0.5  \\
10	&100.0	&99.99±$(<10^{-5})$	&1.0	&0.8042  \\
11	&99.30	&99.30±0.0010	&$<0.01$	&0.3967  \\
12	&91.65	&91.69±0.0026	&$<0.01$	&0.6794  \\
13	&64.19	&64.29±0.0042	&$<0.01$	&0.1061  \\
14	&29.87	&29.93±0.0038	&$<0.01$	&0.0510  \\
15	&9.98	&10.01±0.0026	&$<0.01$	&0.1582  \\
16	&2.81	&2.82±0.0016	&$<0.01$	&0.1172  \\
17	&0.73	&0.74±0.0008	&$<0.01$	&0.1540  \\
18	&0.19	&0.19±0.0004	&0.0110	&0.9427  \\
\hline
 \end{tabular}}

\vspace{.5cm}

\centerline{
\begin{tabular}{ |c | c | c | c | c |}
 \hline
  $\ell$  & Normal & Mela, $\rho=0.0005$& P-value& P-value random \\
    \hline
9	&100.0	&99.99±$(<10^{-5})$	&1.0	&0.5  \\
10	&100.0	&99.99±$(<10^{-5})$	&1.0	&0.5910  \\
11	&99.30	&99.29±0.0010	&0.9987	&0.9998  \\
12	&91.65	&91.69±0.0027	&$<0.01$	&0.9877  \\
13	&64.19	&64.29±0.0043	&$<0.01$	&$<0.01$  \\
14	&29.87	&29.94±0.0041	&$<0.01$	&$<0.01$  \\
15	&9.98	&10.02±0.0027	&$<0.01$	&$<0.01$  \\
16	&2.81	&2.82±0.0015	&$<0.01$	&0.0267  \\
17	&0.73	&0.74±0.0009	&$<0.01$	&0.2552  \\
18	&0.19	&0.19±0.0005	&0.0911	&0.9770  \\
\hline
 \end{tabular}}

\vspace{.5cm}

\subsubsection{EBV indication}

\centerline{
\begin{tabular}{ |c | c | c | c | c |}
 \hline
  $\ell$  & Normal & Lung, $\rho=0.0005$& P-value& P-value random \\
    \hline
9	&100.0	&99.99±$(<10^{-5})$	&1.0	&0.9525  \\
10	&99.94	&99.94±$(<10^{-5})$	&$<0.01$	&0.9863  \\
11	&98.26	&98.29±0.0004	&$<0.01$	&1.0  \\
12	&86.74	&86.82±0.0008	&$<0.01$	&1.0  \\
13	&58.29	&58.39±0.0011	&$<0.01$	&1.0  \\
14	&27.43	&27.49±0.0010	&$<0.01$	&1.0  \\
15	&10.05	&10.07±0.0008	&$<0.01$	&1.0  \\
16	&3.19	&3.20±0.0005	&$<0.01$	&0.9999  \\
17	&1.02	&1.02±0.0003	&$<0.01$	&0.9999  \\
18	&0.33	&0.33±0.0002	&$<0.01$	&0.9997  \\
\hline
 \end{tabular}}

\vspace{.5cm}

\centerline{
\begin{tabular}{ |c | c | c | c | c |}
 \hline
  $\ell$  & Normal & Bladder, $\rho=0.0005$& P-value& P-value random \\
    \hline
9	&100.0	&99.99±$(<10^{-5})$	&1.0	&0.7931  \\
10	&99.94	&99.94±$(<10^{-5})$	&$<0.01$	&0.2122  \\
11	&98.26	&98.29±0.0004	&$<0.01$	&0.9999  \\
12	&86.74	&86.83±0.0009	&$<0.01$	&1.0  \\
13	&58.29	&58.40±0.0011	&$<0.01$	&1.0  \\
14	&27.43	&27.49±0.0010	&$<0.01$	&1.0  \\
15	&10.05	&10.07±0.0008	&$<0.01$	&1.0  \\
16	&3.19	&3.20±0.0006	&$<0.01$	&0.9999  \\
17	&1.02	&1.02±0.0003	&$<0.01$	&0.9999  \\
18	&0.33	&0.33±0.0002	&$<0.01$	&0.9997  \\
\hline
 \end{tabular}}

\vspace{.5cm}

\centerline{
\begin{tabular}{ |c | c | c | c | c |}
 \hline
  $\ell$  & Normal & Mela, $\rho=0.0005$& P-value& P-value random \\
    \hline
9	&100.0	&99.99±$(<10^{-5})$	&1.0	&0.2818  \\
10	&99.94	&99.94±$(<10^{-5})$	&$<0.01$	&0.9409  \\
11	&98.26	&98.28±0.0004	&$<0.01$	&1.0  \\
12	&86.74	&86.82±0.0008	&$<0.01$	&1.0  \\
13	&58.29	&58.38±0.0011	&$<0.01$	&1.0  \\
14	&27.43	&27.48±0.0010	&$<0.01$	&1.0  \\
15	&10.05	&10.06±0.0008	&$<0.01$	&1.0  \\
16	&3.19	&3.20±0.0005	&$<0.01$	&1.0  \\
17	&1.02	&1.02±0.0003	&$<0.01$	&1.0  \\
18	&0.33	&0.33±0.0002	&0.8882	&0.9999  \\
\hline
 \end{tabular}}

\vspace{.5cm}

\subsubsection{HHV indication}

\centerline{
\begin{tabular}{ |c | c | c | c | c |}
 \hline
  $\ell$  & Normal & Lung, $\rho=0.0005$& P-value& P-value random \\
    \hline
9	&100.0	&99.99±$(<10^{-5})$	&1.0	&0.5  \\
10	&99.83	&99.83±0.0001	&$<0.01$	&0.9998  \\
11	&95.96	&96.02±0.0006	&$<0.01$	&1.0  \\
12	&79.24	&79.36±0.0013	&$<0.01$	&1.0  \\
13	&49.20	&49.32±0.0015	&$<0.01$	&$<0.01$  \\
14	&21.84	&21.91±0.0011	&$<0.01$	&$<0.01$  \\
15	&7.91	&7.94±0.0007	&$<0.01$	&$<0.01$  \\
16	&2.73	&2.74±0.0004	&$<0.01$	&$<0.01$  \\
17	&1.05	&1.05±0.0003	&$<0.01$	&$<0.01$  \\
18	&0.48	&0.48±0.0003	&$<0.01$	&$<0.01$  \\
\hline
 \end{tabular}}

\vspace{.5cm}

\centerline{
\begin{tabular}{ |c | c | c | c | c |}
 \hline
  $\ell$  & Normal & Bladder, $\rho=0.0005$& P-value& P-value random \\
    \hline
9	&100.0	&99.99±$(<10^{-5})$	&1.0	&0.0510  \\
10	&99.83	&99.83±0.0001	&$<0.01$	&0.9294  \\
11	&95.96	&96.02±0.0006	&$<0.01$	&0.9999  \\
12	&79.24	&79.37±0.0014	&$<0.01$	&0.9773  \\
13	&49.20	&49.32±0.0015	&$<0.01$	&$<0.01$  \\
14	&21.84	&21.91±0.0012	&$<0.01$	&$<0.01$  \\
15	&7.91	&7.94±0.0007	&$<0.01$	&$<0.01$  \\
16	&2.73	&2.74±0.0005	&$<0.01$	&$<0.01$  \\
17	&1.05	&1.05±0.0003	&$<0.01$	&$<0.01$  \\
18	&0.48	&0.48±0.0002	&$<0.01$	&$<0.01$  \\
\hline
 \end{tabular}}

\vspace{.5cm}

\centerline{
\begin{tabular}{ |c | c | c | c | c |}
 \hline
  $\ell$  & Normal & Mela, $\rho=0.0005$& P-value& P-value random \\
    \hline
9	&100.0	&99.99±$(<10^{-5})$	&1.0	&0.7979  \\
10	&99.83	&99.83±0.0001	&$<0.01$	&0.9999  \\
11	&95.96	&96.01±0.0006	&$<0.01$	&1.0  \\
12	&79.24	&79.36±0.0015	&$<0.01$	&1.0  \\
13	&49.20	&49.32±0.0015	&$<0.01$	&$<0.01$  \\
14	&21.84	&21.91±0.0013	&$<0.01$	&$<0.01$  \\
15	&7.91	&7.94±0.0007	&$<0.01$	&$<0.01$  \\
16	&2.73	&2.74±0.0005	&$<0.01$	&$<0.01$  \\
17	&1.05	&1.05±0.0003	&$<0.01$	&$<0.01$  \\
18	&0.48	&0.48±0.0004	&0.0614	&0.0148  \\
\hline
 \end{tabular}}

\vspace{.5cm}

\subsubsection{HPV indication}

\centerline{
\begin{tabular}{ |c | c | c | c | c |}
 \hline
  $\ell$  & Normal & Lung, $\rho=0.0005$& P-value& P-value random \\
    \hline
9	&100.0	&99.98±0.0	&1.0	&nan  \\
10	&99.97	&99.96±0.0002	&1.0	&0.3552  \\
11	&99.05	&99.04±0.0012	&0.9761	&0.9939  \\
12	&90.48	&90.54±0.0032	&$<0.01$	&0.0528  \\
13	&61.24	&61.37±0.0048	&$<0.01$	&$<0.01$  \\
14	&28.22	&28.30±0.0042	&$<0.01$	&$<0.01$  \\
15	&9.84	&9.88±0.0030	&$<0.01$	&$<0.01$  \\
16	&3.19	&3.20±0.0017	&$<0.01$	&0.0111  \\
17	&1.10	&1.10±0.0010	&$<0.01$	&0.4128  \\
18	&0.50	&0.50±0.0006	&$<0.01$	&0.5864  \\
\hline
 \end{tabular}}

\vspace{.5cm}

\centerline{
\begin{tabular}{ |c | c | c | c | c |}
 \hline
  $\ell$  & Normal & Bladder, $\rho=0.0005$& P-value& P-value random \\
    \hline
9	&100.0	&99.98±0.0	&1.0	&nan  \\
10	&99.97	&99.96±0.0002	&1.0	&0.8341  \\
11	&99.05	&99.04±0.0011	&0.9898	&0.9969  \\
12	&90.48	&90.54±0.0030	&$<0.01$	&0.5926  \\
13	&61.24	&61.36±0.0049	&$<0.01$	&$<0.01$  \\
14	&28.22	&28.29±0.0040	&$<0.01$	&$<0.01$  \\
15	&9.84	&9.88±0.0029	&$<0.01$	&$<0.01$  \\
16	&3.19	&3.20±0.0017	&$<0.01$	&0.0160  \\
17	&1.10	&1.10±0.0010	&$<0.01$	&0.8741  \\
18	&0.50	&0.50±0.0005	&$<0.01$	&0.7670  \\
\hline
 \end{tabular}}

\vspace{.5cm}

\centerline{
\begin{tabular}{ |c | c | c | c | c |}
 \hline
  $\ell$  & Normal & Mela, $\rho=0.0005$& P-value& P-value random \\
    \hline
9	&100.0	&99.98±0.0	&1.0	&nan  \\
10	&99.97	&99.96±0.0003	&1.0	&0.2953  \\
11	&99.05	&99.04±0.0011	&0.9943	&0.9980  \\
12	&90.48	&90.54±0.0030	&$<0.01$	&$<0.01$  \\
13	&61.24	&61.38±0.0046	&$<0.01$	&$<0.01$  \\
14	&28.22	&28.31±0.0041	&$<0.01$	&$<0.01$  \\
15	&9.84	&9.88±0.0030	&$<0.01$	&$<0.01$  \\
16	&3.19	&3.20±0.0018	&$<0.01$	&$<0.01$  \\
17	&1.10	&1.10±0.0010	&$<0.01$	&0.3923  \\
18	&0.50	&0.50±0.0006	&$<0.01$	&0.6700  \\
\hline
 \end{tabular}}

\vspace{.5cm}

\subsubsection{Measles virus indication}

\centerline{
\begin{tabular}{ |c | c | c | c | c |}
 \hline
  $\ell$  & Normal & Lung, $\rho=0.0005$& P-value& P-value random \\
    \hline
9	&100.0	&99.99±$(<10^{-5})$	&1.0	&0.5  \\
10	&99.97	&99.96±0.0002	&1.0	&0.9512  \\
11	&98.23	&98.25±0.0012	&$<0.01$	&0.9931  \\
12	&86.15	&86.24±0.0025	&$<0.01$	&0.0257  \\
13	&54.27	&54.40±0.0035	&$<0.01$	&$<0.01$  \\
14	&22.64	&22.72±0.0030	&$<0.01$	&$<0.01$  \\
15	&7.29	&7.32±0.0019	&$<0.01$	&$<0.01$  \\
16	&2.19	&2.20±0.0011	&$<0.01$	&0.0878  \\
17	&0.68	&0.68±0.0006	&0.7775	&0.9909  \\
18	&0.19	&0.19±0.0003	&0.1692	&0.4103  \\
\hline
 \end{tabular}}

\vspace{.5cm}

\centerline{
\begin{tabular}{ |c | c | c | c | c |}
 \hline
  $\ell$  & Normal & Bladder, $\rho=0.0005$& P-value& P-value random \\
    \hline
9	&100.0	&99.99±$(<10^{-5})$	&1.0	&0.5  \\
10	&99.97	&99.96±0.0002	&1.0	&0.8931  \\
11	&98.23	&98.25±0.0012	&$<0.01$	&0.9887  \\
12	&86.15	&86.24±0.0027	&$<0.01$	&0.6929  \\
13	&54.27	&54.40±0.0034	&$<0.01$	&0.0118  \\
14	&22.64	&22.72±0.0029	&$<0.01$	&$<0.01$  \\
15	&7.29	&7.32±0.0020	&$<0.01$	&0.3013  \\
16	&2.19	&2.20±0.0011	&$<0.01$	&0.4939  \\
17	&0.68	&0.68±0.0006	&0.0737	&0.8014  \\
18	&0.19	&0.19±0.0003	&0.6188	&0.7503  \\
\hline
 \end{tabular}}

\vspace{.5cm}

\centerline{
\begin{tabular}{ |c | c | c | c | c |}
 \hline
  $\ell$  & Normal & Mela, $\rho=0.0005$& P-value& P-value random \\
    \hline
9	&100.0	&99.99±$(<10^{-5})$	&1.0	&0.5  \\
10	&99.97	&99.96±0.0002	&1.0	&0.9656  \\
11	&98.23	&98.25±0.0012	&$<0.01$	&0.9999  \\
12	&86.15	&86.24±0.0025	&$<0.01$	&0.5577  \\
13	&54.27	&54.40±0.0034	&$<0.01$	&$<0.01$  \\
14	&22.64	&22.72±0.0030	&$<0.01$	&$<0.01$  \\
15	&7.29	&7.32±0.0020	&$<0.01$	&0.1858  \\
16	&2.19	&2.20±0.0011	&$<0.01$	&0.4911  \\
17	&0.68	&0.68±0.0006	&0.2713	&0.9212  \\
18	&0.19	&0.19±0.0003	&0.9773	&0.9687  \\
\hline
 \end{tabular}}

\vspace{.5cm}

\subsubsection{Yellow fever virus indication}

\centerline{
\begin{tabular}{ |c | c | c | c | c |}
 \hline
  $\ell$  & Normal & Lung, $\rho=0.0005$& P-value& P-value random \\
    \hline
9	&100.0	&99.99±$(<10^{-5})$	&1.0	&0.5  \\
10	&100.0	&99.98±0.0001	&1.0	&0.7992  \\
11	&99.05	&99.05±0.0011	&$<0.01$	&0.9959  \\
12	&92.10	&92.16±0.0028	&$<0.01$	&0.7110  \\
13	&65.28	&65.38±0.0042	&$<0.01$	&$<0.01$  \\
14	&30.54	&30.61±0.0040	&$<0.01$	&$<0.01$  \\
15	&10.81	&10.83±0.0028	&$<0.01$	&$<0.01$  \\
16	&3.15	&3.16±0.0017	&$<0.01$	&0.3423  \\
17	&0.87	&0.87±0.0009	&$<0.01$	&0.0814  \\
18	&0.29	&0.29±0.0005	&$<0.01$	&0.1116  \\
\hline
 \end{tabular}}

\vspace{.5cm}

\centerline{
\begin{tabular}{ |c | c | c | c | c |}
 \hline
  $\ell$  & Normal & Bladder, $\rho=0.0005$& P-value& P-value random \\
    \hline
9	&100.0	&99.99±$(<10^{-5})$	&1.0	&0.5  \\
10	&100.0	&99.98±0.0001	&1.0	&0.4066  \\
11	&99.05	&99.05±0.0011	&$<0.01$	&0.9801  \\
12	&92.10	&92.16±0.0027	&$<0.01$	&0.7498  \\
13	&65.28	&65.38±0.0041	&$<0.01$	&0.0445  \\
14	&30.54	&30.61±0.0042	&$<0.01$	&0.0112  \\
15	&10.81	&10.83±0.0028	&$<0.01$	&0.2486  \\
16	&3.15	&3.16±0.0017	&$<0.01$	&0.6824  \\
17	&0.87	&0.87±0.0009	&$<0.01$	&0.0726  \\
18	&0.29	&0.29±0.0006	&$<0.01$	&$<0.01$  \\
\hline
 \end{tabular}}

\vspace{.5cm}

\centerline{
\begin{tabular}{ |c | c | c | c | c |}
 \hline
  $\ell$  & Normal & Mela, $\rho=0.0005$& P-value& P-value random \\
    \hline
9	&100.0	&99.99±$(<10^{-5})$	&1.0	&0.5  \\
10	&100.0	&99.98±0.0001	&1.0	&0.3586  \\
11	&99.05	&99.05±0.0011	&$<0.01$	&0.9999  \\
12	&92.10	&92.16±0.0025	&$<0.01$	&0.9998  \\
13	&65.28	&65.38±0.0043	&$<0.01$	&0.0127  \\
14	&30.54	&30.61±0.0040	&$<0.01$	&$<0.01$  \\
15	&10.81	&10.83±0.0028	&$<0.01$	&0.0410  \\
16	&3.15	&3.16±0.0016	&$<0.01$	&0.2823  \\
17	&0.87	&0.87±0.0009	&$<0.01$	&0.2685  \\
18	&0.29	&0.29±0.0005	&$<0.01$	&0.0670  \\
\hline
 \end{tabular}}

\vspace{.5cm}

\subsection{$\rho=0.001$}\label{r2}
\subsubsection{Ebola virus indication}

\centerline{
\begin{tabular}{ |c | c | c | c | c |}
 \hline
  $\ell$  & Normal & Lung, $\rho=0.001$& P-value& P-value random \\
    \hline
9	&100.0	&99.99±$(<10^{-5})$	&1.0	&0.5  \\
10	&99.94	&99.94±0.0003	&1.0	&0.3328  \\
11	&98.11	&98.16±0.0015	&$<0.01$	&0.9999  \\
12	&86.99	&87.16±0.0031	&$<0.01$	&0.2037  \\
13	&56.32	&56.58±0.0045	&$<0.01$	&$<0.01$  \\
14	&23.94	&24.11±0.0040	&$<0.01$	&$<0.01$  \\
15	&7.82	&7.88±0.0025	&$<0.01$	&$<0.01$  \\
16	&2.40	&2.42±0.0014	&$<0.01$	&$<0.01$  \\
17	&0.62	&0.63±0.0008	&$<0.01$	&$<0.01$  \\
18	&0.12	&0.12±0.0004	&$<0.01$	&0.0393  \\
\hline
 \end{tabular}}

\vspace{.5cm}

\centerline{
\begin{tabular}{ |c | c | c | c | c |}
 \hline
  $\ell$  & Normal & Bladder, $\rho=0.001$& P-value& P-value random \\
    \hline
9	&100.0	&99.99±$(<10^{-5})$	&1.0	&0.5  \\
10	&99.94	&99.94±0.0003	&0.9999	&$<0.01$  \\
11	&98.11	&98.17±0.0015	&$<0.01$	&0.9999  \\
12	&86.99	&87.16±0.0031	&$<0.01$	&0.4714  \\
13	&56.32	&56.57±0.0043	&$<0.01$	&$<0.01$  \\
14	&23.94	&24.10±0.0038	&$<0.01$	&$<0.01$  \\
15	&7.82	&7.88±0.0025	&$<0.01$	&$<0.01$  \\
16	&2.40	&2.42±0.0014	&$<0.01$	&$<0.01$  \\
17	&0.62	&0.63±0.0008	&$<0.01$	&$<0.01$  \\
18	&0.12	&0.12±0.0004	&$<0.01$	&0.0521  \\
\hline
 \end{tabular}}

\vspace{.5cm}

\centerline{
\begin{tabular}{ |c | c | c | c | c |}
 \hline
  $\ell$  & Normal & Mela, $\rho=0.001$& P-value& P-value random \\
    \hline
9	&100.0	&99.99±$(<10^{-5})$	&1.0	&0.5  \\
10	&99.94	&99.94±0.0002	&1.0	&0.8304  \\
11	&98.11	&98.16±0.0015	&$<0.01$	&1.0  \\
12	&86.99	&87.15±0.0032	&$<0.01$	&0.9989  \\
13	&56.32	&56.59±0.0045	&$<0.01$	&$<0.01$  \\
14	&23.94	&24.12±0.0039	&$<0.01$	&$<0.01$  \\
15	&7.82	&7.89±0.0025	&$<0.01$	&$<0.01$  \\
16	&2.40	&2.42±0.0014	&$<0.01$	&$<0.01$  \\
17	&0.62	&0.63±0.0007	&$<0.01$	&$<0.01$  \\
18	&0.12	&0.12±0.0004	&$<0.01$	&0.1217  \\
\hline
 \end{tabular}}

\vspace{.5cm}

\subsubsection{CMV indication}

\centerline{
\begin{tabular}{ |c | c | c | c | c |}
 \hline
  $\ell$  & Normal & Lung, $\rho=0.001$& P-value& P-value random \\
    \hline
9	&100.0	&99.99±$(<10^{-5})$	&1.0	&0.6082  \\
10	&99.71	&99.73±0.0002	&$<0.01$	&0.9999  \\
11	&94.88	&95.01±0.0007	&$<0.01$	&1.0  \\
12	&74.92	&75.18±0.0011	&$<0.01$	&1.0  \\
13	&43.52	&43.74±0.0012	&$<0.01$	&1.0  \\
14	&18.64	&18.75±0.0010	&$<0.01$	&1.0  \\
15	&6.60	&6.63±0.0006	&$<0.01$	&1.0  \\
16	&2.23	&2.24±0.0003	&$<0.01$	&1.0  \\
17	&0.77	&0.77±0.0002	&$<0.01$	&0.9999  \\
18	&0.27	&0.27±0.0001	&$<0.01$	&0.9987  \\
\hline
 \end{tabular}}

\vspace{.5cm}

\centerline{
\begin{tabular}{ |c | c | c | c | c |}
 \hline
  $\ell$  & Normal & Bladder, $\rho=0.001$& P-value& P-value random \\
    \hline
9	&100.0	&99.99±$(<10^{-5})$	&1.0	&0.5183  \\
10	&99.71	&99.73±0.0002	&$<0.01$	&0.0255  \\
11	&94.88	&95.02±0.0007	&$<0.01$	&0.9999  \\
12	&74.92	&75.21±0.0011	&$<0.01$	&1.0  \\
13	&43.52	&43.76±0.0012	&$<0.01$	&1.0  \\
14	&18.64	&18.76±0.0010	&$<0.01$	&1.0  \\
15	&6.60	&6.64±0.0007	&$<0.01$	&0.9999  \\
16	&2.23	&2.24±0.0004	&$<0.01$	&0.9999  \\
17	&0.77	&0.77±0.0002	&$<0.01$	&0.5676  \\
18	&0.27	&0.27±0.0001	&$<0.01$	&0.3313  \\
\hline
 \end{tabular}}

\vspace{.5cm}

\centerline{
\begin{tabular}{ |c | c | c | c | c |}
 \hline
  $\ell$  & Normal & Mela, $\rho=0.001$& P-value& P-value random \\
    \hline
9	&100.0	&99.99±$(<10^{-5})$	&1.0	&0.7682  \\
10	&99.71	&99.73±0.0002	&$<0.01$	&0.9999  \\
11	&94.88	&95.00±0.0007	&$<0.01$	&1.0  \\
12	&74.92	&75.16±0.0011	&$<0.01$	&1.0  \\
13	&43.52	&43.72±0.0012	&$<0.01$	&1.0  \\
14	&18.64	&18.73±0.0010	&$<0.01$	&1.0  \\
15	&6.60	&6.63±0.0006	&$<0.01$	&1.0  \\
16	&2.23	&2.24±0.0004	&$<0.01$	&1.0  \\
17	&0.77	&0.77±0.0002	&$<0.01$	&0.9999  \\
18	&0.27	&0.27±0.0001	&$<0.01$	&0.7692  \\
\hline
 \end{tabular}}

\vspace{.5cm}

\subsubsection{Dengue virus indication}

\centerline{
\begin{tabular}{ |c | c | c | c | c |}
 \hline
  $\ell$  & Normal & Lung, $\rho=0.001$& P-value& P-value random \\
    \hline
9	&100.0	&99.99±$(<10^{-5})$	&1.0	&0.5  \\
10	&100.0	&99.98±$(<10^{-5})$	&1.0	&0.3914  \\
11	&99.30	&99.31±0.0014	&$<0.01$	&0.9896  \\
12	&91.65	&91.75±0.0038	&$<0.01$	&0.9918  \\
13	&64.19	&64.40±0.0060	&$<0.01$	&$<0.01$  \\
14	&29.87	&30.01±0.0056	&$<0.01$	&$<0.01$  \\
15	&9.98	&10.04±0.0039	&$<0.01$	&$<0.01$  \\
16	&2.81	&2.83±0.0023	&$<0.01$	&$<0.01$  \\
17	&0.73	&0.74±0.0012	&$<0.01$	&0.4424  \\
18	&0.19	&0.19±0.0005	&0.0456	&0.9791  \\
\hline
 \end{tabular}}

\vspace{.5cm}

\centerline{
\begin{tabular}{ |c | c | c | c | c |}
 \hline
  $\ell$  & Normal & Bladder, $\rho=0.001$& P-value& P-value random \\
    \hline
9	&100.0	&99.99±$(<10^{-5})$	&1.0	&0.5  \\
10	&100.0	&99.98±0.0001	&1.0	&0.8145  \\
11	&99.30	&99.31±0.0014	&$<0.01$	&0.9694  \\
12	&91.65	&91.75±0.0036	&$<0.01$	&0.9986  \\
13	&64.19	&64.39±0.0059	&$<0.01$	&$<0.01$  \\
14	&29.87	&30.01±0.0057	&$<0.01$	&$<0.01$  \\
15	&9.98	&10.04±0.0039	&$<0.01$	&$<0.01$  \\
16	&2.81	&2.83±0.0021	&$<0.01$	&0.0433  \\
17	&0.73	&0.74±0.0012	&$<0.01$	&0.5961  \\
18	&0.19	&0.19±0.0006	&$<0.01$	&0.8638  \\
\hline
 \end{tabular}}

\vspace{.5cm}

\centerline{
\begin{tabular}{ |c | c | c | c | c |}
 \hline
  $\ell$  & Normal & Mela, $\rho=0.001$& P-value& P-value random \\
    \hline
9	&100.0	&99.99±$(<10^{-5})$	&1.0	&0.5  \\
10	&100.0	&99.98±$(<10^{-5})$	&1.0	&0.4457  \\
11	&99.30	&99.30±0.0013	&$<0.01$	&0.9999  \\
12	&91.65	&91.75±0.0035	&$<0.01$	&0.9999  \\
13	&64.19	&64.41±0.0061	&$<0.01$	&$<0.01$  \\
14	&29.87	&30.02±0.0055	&$<0.01$	&$<0.01$  \\
15	&9.98	&10.05±0.0037	&$<0.01$	&$<0.01$  \\
16	&2.81	&2.83±0.0023	&$<0.01$	&0.2108  \\
17	&0.73	&0.74±0.0012	&$<0.01$	&0.3675  \\
18	&0.19	&0.19±0.0006	&$<0.01$	&0.7893  \\
\hline
 \end{tabular}}

\vspace{.5cm}

\subsubsection{EBV indication}

\centerline{
\begin{tabular}{ |c | c | c | c | c |}
 \hline
  $\ell$  & Normal & Lung, $\rho=0.001$& P-value& P-value random \\
    \hline
9	&100.0	&99.99±$(<10^{-5})$	&1.0	&0.1418  \\
10	&99.94	&99.95±0.0001	&$<0.01$	&0.7993  \\
11	&98.26	&98.31±0.0005	&$<0.01$	&1.0  \\
12	&86.74	&86.91±0.0012	&$<0.01$	&1.0  \\
13	&58.29	&58.49±0.0016	&$<0.01$	&1.0  \\
14	&27.43	&27.54±0.0015	&$<0.01$	&1.0  \\
15	&10.05	&10.08±0.0012	&$<0.01$	&1.0  \\
16	&3.19	&3.20±0.0008	&$<0.01$	&1.0  \\
17	&1.02	&1.03±0.0004	&$<0.01$	&0.9999  \\
18	&0.33	&0.33±0.0002	&$<0.01$	&0.9999  \\
\hline
 \end{tabular}}

\vspace{.5cm}

\centerline{
\begin{tabular}{ |c | c | c | c | c |}
 \hline
  $\ell$  & Normal & Bladder, $\rho=0.001$& P-value& P-value random \\
    \hline
9	&100.0	&99.99±$(<10^{-5})$	&1.0	&0.2185  \\
10	&99.94	&99.95±0.0001	&$<0.01$	&0.1288  \\
11	&98.26	&98.31±0.0005	&$<0.01$	&0.9999  \\
12	&86.74	&86.92±0.0012	&$<0.01$	&1.0  \\
13	&58.29	&58.50±0.0016	&$<0.01$	&1.0  \\
14	&27.43	&27.55±0.0014	&$<0.01$	&1.0  \\
15	&10.05	&10.09±0.0012	&$<0.01$	&1.0  \\
16	&3.19	&3.20±0.0008	&$<0.01$	&1.0  \\
17	&1.02	&1.03±0.0004	&$<0.01$	&0.9999  \\
18	&0.33	&0.33±0.0003	&$<0.01$	&0.9999  \\
\hline
 \end{tabular}}

\vspace{.5cm}

\centerline{
\begin{tabular}{ |c | c | c | c | c |}
 \hline
  $\ell$  & Normal & Mela, $\rho=0.001$& P-value& P-value random \\
    \hline
9	&100.0	&99.99±$(<10^{-5})$	&1.0	&0.0412  \\
10	&99.94	&99.95±0.0001	&$<0.01$	&0.9878  \\
11	&98.26	&98.31±0.0005	&$<0.01$	&1.0  \\
12	&86.74	&86.90±0.0012	&$<0.01$	&1.0  \\
13	&58.29	&58.47±0.0016	&$<0.01$	&1.0  \\
14	&27.43	&27.52±0.0014	&$<0.01$	&1.0  \\
15	&10.05	&10.07±0.0011	&$<0.01$	&1.0  \\
16	&3.19	&3.20±0.0008	&$<0.01$	&1.0  \\
17	&1.02	&1.02±0.0004	&$<0.01$	&1.0  \\
18	&0.33	&0.33±0.0002	&0.9238	&1.0  \\
\hline
 \end{tabular}}

\vspace{.5cm}

\subsubsection{HHV indication}

\centerline{
\begin{tabular}{ |c | c | c | c | c |}
 \hline
  $\ell$  & Normal & Lung, $\rho=0.001$& P-value& P-value random \\
    \hline
9	&100.0	&99.99±$(<10^{-5})$	&1.0	&0.2959  \\
10	&99.83	&99.84±0.0002	&$<0.01$	&1.0  \\
11	&95.96	&96.07±0.0008	&$<0.01$	&1.0  \\
12	&79.24	&79.48±0.0018	&$<0.01$	&1.0  \\
13	&49.20	&49.45±0.0021	&$<0.01$	&$<0.01$  \\
14	&21.84	&21.98±0.0018	&$<0.01$	&$<0.01$  \\
15	&7.91	&7.97±0.0010	&$<0.01$	&$<0.01$  \\
16	&2.73	&2.75±0.0007	&$<0.01$	&$<0.01$  \\
17	&1.05	&1.06±0.0005	&$<0.01$	&$<0.01$  \\
18	&0.48	&0.48±0.0004	&$<0.01$	&$<0.01$  \\
\hline
 \end{tabular}}

\vspace{.5cm}

\centerline{
\begin{tabular}{ |c | c | c | c | c |}
 \hline
  $\ell$  & Normal & Bladder, $\rho=0.001$& P-value& P-value random \\
    \hline
9	&100.0	&99.99±$(<10^{-5})$	&1.0	&0.5000  \\
10	&99.83	&99.84±0.0002	&$<0.01$	&0.9739  \\
11	&95.96	&96.08±0.0008	&$<0.01$	&1.0  \\
12	&79.24	&79.49±0.0018	&$<0.01$	&0.9999  \\
13	&49.20	&49.45±0.0023	&$<0.01$	&$<0.01$  \\
14	&21.84	&21.98±0.0018	&$<0.01$	&$<0.01$  \\
15	&7.91	&7.97±0.0010	&$<0.01$	&$<0.01$  \\
16	&2.73	&2.75±0.0007	&$<0.01$	&$<0.01$  \\
17	&1.05	&1.06±0.0004	&$<0.01$	&$<0.01$  \\
18	&0.48	&0.48±0.0005	&$<0.01$	&$<0.01$  \\
\hline
 \end{tabular}}

\vspace{.5cm}

\centerline{
\begin{tabular}{ |c | c | c | c | c |}
 \hline
  $\ell$  & Normal & Mela, $\rho=0.001$& P-value& P-value random \\
    \hline
9	&100.0	&99.99±$(<10^{-5})$	&1.0	&0.8155  \\
10	&99.83	&99.84±0.0002	&$<0.01$	&1.0  \\
11	&95.96	&96.06±0.0008	&$<0.01$	&1.0  \\
12	&79.24	&79.47±0.0020	&$<0.01$	&1.0  \\
13	&49.20	&49.45±0.0022	&$<0.01$	&$<0.01$  \\
14	&21.84	&21.99±0.0017	&$<0.01$	&$<0.01$  \\
15	&7.91	&7.97±0.0011	&$<0.01$	&$<0.01$  \\
16	&2.73	&2.75±0.0007	&$<0.01$	&$<0.01$  \\
17	&1.05	&1.06±0.0004	&$<0.01$	&$<0.01$  \\
18	&0.48	&0.48±0.0005	&$<0.01$	&$<0.01$  \\
\hline
 \end{tabular}}

\vspace{.5cm}

\subsubsection{HPV indication}

\centerline{
\begin{tabular}{ |c | c | c | c | c |}
 \hline
  $\ell$  & Normal & Lung, $\rho=0.001$& P-value& P-value random \\
    \hline
9	&100.0	&99.98±$(<10^{-5})$	&1.0	&0.5  \\
10	&99.97	&99.96±0.0004	&1.0	&0.4823  \\
11	&99.05	&99.05±0.0016	&$<0.01$	&0.9992  \\
12	&90.48	&90.61±0.0044	&$<0.01$	&0.0447  \\
13	&61.24	&61.48±0.0068	&$<0.01$	&$<0.01$  \\
14	&28.22	&28.39±0.0062	&$<0.01$	&$<0.01$  \\
15	&9.84	&9.91±0.0042	&$<0.01$	&$<0.01$  \\
16	&3.19	&3.22±0.0024	&$<0.01$	&$<0.01$  \\
17	&1.10	&1.11±0.0015	&$<0.01$	&0.0765  \\
18	&0.50	&0.50±0.0009	&$<0.01$	&0.3499  \\
\hline
 \end{tabular}}

\vspace{.5cm}

\centerline{
\begin{tabular}{ |c | c | c | c | c |}
 \hline
  $\ell$  & Normal & Bladder, $\rho=0.001$& P-value& P-value random \\
    \hline
9	&100.0	&99.98±$(<10^{-5})$	&1.0	&0.5  \\
10	&99.97	&99.96±0.0003	&1.0	&0.2536  \\
11	&99.05	&99.06±0.0016	&$<0.01$	&0.9969  \\
12	&90.48	&90.61±0.0044	&$<0.01$	&0.2479  \\
13	&61.24	&61.47±0.0067	&$<0.01$	&$<0.01$  \\
14	&28.22	&28.38±0.0061	&$<0.01$	&$<0.01$  \\
15	&9.84	&9.91±0.0040	&$<0.01$	&$<0.01$  \\
16	&3.19	&3.21±0.0025	&$<0.01$	&$<0.01$  \\
17	&1.10	&1.10±0.0014	&$<0.01$	&0.4720  \\
18	&0.50	&0.50±0.0008	&$<0.01$	&0.8303  \\
\hline
 \end{tabular}}

\vspace{.5cm}

\centerline{
\begin{tabular}{ |c | c | c | c | c |}
 \hline
  $\ell$  & Normal & Mela, $\rho=0.001$& P-value& P-value random \\
    \hline
9	&100.0	&99.98±$(<10^{-5})$	&1.0	&0.5  \\
10	&99.97	&99.96±0.0003	&1.0	&0.4093  \\
11	&99.05	&99.05±0.0016	&$<0.01$	&0.9998  \\
12	&90.48	&90.62±0.0043	&$<0.01$	&$<0.01$  \\
13	&61.24	&61.50±0.0069	&$<0.01$	&$<0.01$  \\
14	&28.22	&28.40±0.0059	&$<0.01$	&$<0.01$  \\
15	&9.84	&9.92±0.0040	&$<0.01$	&$<0.01$  \\
16	&3.19	&3.22±0.0025	&$<0.01$	&$<0.01$  \\
17	&1.10	&1.11±0.0014	&$<0.01$	&0.0188  \\
18	&0.50	&0.50±0.0008	&$<0.01$	&0.3983  \\
\hline
 \end{tabular}}

\vspace{.5cm}

\subsubsection{Measles virus indication}

\centerline{
\begin{tabular}{ |c | c | c | c | c |}
 \hline
  $\ell$  & Normal & Lung, $\rho=0.001$& P-value& P-value random \\
    \hline
9	&100.0	&99.99±$(<10^{-5})$	&1.0	&0.5  \\
10	&99.97	&99.97±0.0003	&1.0	&0.4318  \\
11	&98.23	&98.28±0.0017	&$<0.01$	&0.9998  \\
12	&86.15	&86.34±0.0036	&$<0.01$	&0.5411  \\
13	&54.27	&54.53±0.0050	&$<0.01$	&$<0.01$  \\
14	&22.64	&22.79±0.0043	&$<0.01$	&$<0.01$  \\
15	&7.29	&7.34±0.0028	&$<0.01$	&0.1535  \\
16	&2.19	&2.20±0.0015	&$<0.01$	&0.3683  \\
17	&0.68	&0.68±0.0008	&0.0104	&0.4496  \\
18	&0.19	&0.19±0.0004	&0.9896	&0.7992  \\
\hline
 \end{tabular}}

\vspace{.5cm}

\centerline{
\begin{tabular}{ |c | c | c | c | c |}
 \hline
  $\ell$  & Normal & Bladder, $\rho=0.001$& P-value& P-value random \\
    \hline
9	&100.0	&99.99±$(<10^{-5})$	&1.0	&0.5  \\
10	&99.97	&99.97±0.0003	&1.0	&0.3667  \\
11	&98.23	&98.28±0.0017	&$<0.01$	&0.9969  \\
12	&86.15	&86.34±0.0035	&$<0.01$	&0.1666  \\
13	&54.27	&54.52±0.0050	&$<0.01$	&$<0.01$  \\
14	&22.64	&22.78±0.0042	&$<0.01$	&$<0.01$  \\
15	&7.29	&7.34±0.0029	&$<0.01$	&0.9235  \\
16	&2.19	&2.20±0.0016	&$<0.01$	&0.5732  \\
17	&0.68	&0.68±0.0009	&0.2049	&0.7833  \\
18	&0.19	&0.19±0.0004	&0.3444	&0.1426  \\
\hline
 \end{tabular}}

\vspace{.5cm}

\centerline{
\begin{tabular}{ |c | c | c | c | c |}
 \hline
  $\ell$  & Normal & Mela, $\rho=0.001$& P-value& P-value random \\
    \hline
9	&100.0	&99.99±$(<10^{-5})$	&1.0	&0.5  \\
10	&99.97	&99.97±0.0002	&1.0	&0.1908  \\
11	&98.23	&98.27±0.0015	&$<0.01$	&0.9999  \\
12	&86.15	&86.34±0.0037	&$<0.01$	&0.9581  \\
13	&54.27	&54.53±0.0047	&$<0.01$	&$<0.01$  \\
14	&22.64	&22.78±0.0041	&$<0.01$	&$<0.01$  \\
15	&7.29	&7.34±0.0027	&$<0.01$	&0.9399  \\
16	&2.19	&2.20±0.0016	&$<0.01$	&0.2985  \\
17	&0.68	&0.68±0.0009	&0.3218	&0.8489  \\
18	&0.19	&0.19±0.0004	&0.8608	&0.4713  \\
\hline
 \end{tabular}}

\vspace{.5cm}

\subsubsection{Yellow fever virus indication}

\centerline{
\begin{tabular}{ |c | c | c | c | c |}
 \hline
  $\ell$  & Normal & Lung, $\rho=0.001$& P-value& P-value random \\
    \hline
9	&100.0	&99.99±$(<10^{-5})$	&1.0	&0.5  \\
10	&100.0	&99.98±0.0001	&1.0	&0.9805  \\
11	&99.05	&99.07±0.0015	&$<0.01$	&0.9999  \\
12	&92.10	&92.21±0.0038	&$<0.01$	&0.9061  \\
13	&65.28	&65.48±0.0059	&$<0.01$	&0.6362  \\
14	&30.54	&30.68±0.0056	&$<0.01$	&0.1009  \\
15	&10.81	&10.86±0.0040	&$<0.01$	&0.2397  \\
16	&3.15	&3.17±0.0024	&$<0.01$	&0.7796  \\
17	&0.87	&0.87±0.0013	&$<0.01$	&0.0412  \\
18	&0.29	&0.29±0.0007	&$<0.01$	&0.1197  \\
\hline
 \end{tabular}}

\vspace{.5cm}

\centerline{
\begin{tabular}{ |c | c | c | c | c |}
 \hline
  $\ell$  & Normal & Bladder, $\rho=0.001$& P-value& P-value random \\
    \hline
9	&100.0	&99.99±$(<10^{-5})$	&1.0	&0.5  \\
10	&100.0	&99.98±0.0001	&1.0	&0.9826  \\
11	&99.05	&99.07±0.0015	&$<0.01$	&0.8695  \\
12	&92.10	&92.21±0.0038	&$<0.01$	&0.9591  \\
13	&65.28	&65.47±0.0062	&$<0.01$	&0.8768  \\
14	&30.54	&30.67±0.0058	&$<0.01$	&0.6535  \\
15	&10.81	&10.85±0.0040	&$<0.01$	&0.3259  \\
16	&3.15	&3.17±0.0024	&$<0.01$	&0.7424  \\
17	&0.87	&0.88±0.0013	&$<0.01$	&$<0.01$  \\
18	&0.29	&0.29±0.0008	&$<0.01$	&0.0134  \\
\hline
 \end{tabular}}

\vspace{.5cm}

\centerline{
\begin{tabular}{ |c | c | c | c | c |}
 \hline
  $\ell$  & Normal & Mela, $\rho=0.001$& P-value& P-value random \\
    \hline
9	&100.0	&99.99±$(<10^{-5})$	&1.0	&0.5  \\
10	&100.0	&99.98±0.0001	&1.0	&0.9725  \\
11	&99.05	&99.06±0.0015	&$<0.01$	&0.9999  \\
12	&92.10	&92.20±0.0038	&$<0.01$	&0.9993  \\
13	&65.28	&65.48±0.0059	&$<0.01$	&0.4637  \\
14	&30.54	&30.68±0.0057	&$<0.01$	&0.1347  \\
15	&10.81	&10.86±0.0039	&$<0.01$	&0.0967  \\
16	&3.15	&3.17±0.0023	&$<0.01$	&0.0534  \\
17	&0.87	&0.87±0.0014	&$<0.01$	&$<0.01$  \\
18	&0.29	&0.29±0.0007	&$<0.01$	&0.0580  \\
\hline
 \end{tabular}}

\vspace{.5cm}

\subsection{$\rho=0.01$}\label{r3}
\subsubsection{Ebola virus indication}

\centerline{
\begin{tabular}{ |c | c | c | c | c |}
 \hline
  $\ell$  & Normal & Lung, $\rho=0.01$& P-value& P-value random \\
    \hline
9	&100.0	&99.99±$(<10^{-5})$	&1.0	&0.5  \\
10	&99.94	&99.96±0.0005	&$<0.01$	&0.9989  \\
11	&98.11	&98.61±0.0035	&$<0.01$	&1.0  \\
12	&86.99	&88.54±0.0080	&$<0.01$	&0.9074  \\
13	&56.32	&58.74±0.0125	&$<0.01$	&$<0.01$  \\
14	&23.94	&25.48±0.0119	&$<0.01$	&$<0.01$  \\
15	&7.82	&8.41±0.0079	&$<0.01$	&$<0.01$  \\
16	&2.40	&2.57±0.0045	&$<0.01$	&$<0.01$  \\
17	&0.62	&0.68±0.0025	&$<0.01$	&$<0.01$  \\
18	&0.12	&0.14±0.0013	&$<0.01$	&0.0210  \\
\hline
 \end{tabular}}

\vspace{.5cm}

\centerline{
\begin{tabular}{ |c | c | c | c | c |}
 \hline
  $\ell$  & Normal & Bladder, $\rho=0.01$& P-value& P-value random \\
    \hline
9	&100.0	&99.99±$(<10^{-5})$	&1.0	&0.5  \\
10	&99.94	&99.96±0.0005	&$<0.01$	&0.8464  \\
11	&98.11	&98.62±0.0036	&$<0.01$	&1.0  \\
12	&86.99	&88.54±0.0083	&$<0.01$	&0.8328  \\
13	&56.32	&58.63±0.0124	&$<0.01$	&$<0.01$  \\
14	&23.94	&25.40±0.0112	&$<0.01$	&$<0.01$  \\
15	&7.82	&8.38±0.0076	&$<0.01$	&$<0.01$  \\
16	&2.40	&2.56±0.0044	&$<0.01$	&$<0.01$  \\
17	&0.62	&0.68±0.0024	&$<0.01$	&$<0.01$  \\
18	&0.12	&0.14±0.0013	&$<0.01$	&$<0.01$  \\
\hline
 \end{tabular}}

\vspace{.5cm}

\centerline{
\begin{tabular}{ |c | c | c | c | c |}
 \hline
  $\ell$  & Normal & Mela, $\rho=0.01$& P-value& P-value random \\
    \hline
9	&100.0	&99.99±$(<10^{-5})$	&1.0	&1.0  \\
10	&99.94	&99.96±0.0005	&$<0.01$	&0.9999  \\
11	&98.11	&98.55±0.0035	&$<0.01$	&1.0  \\
12	&86.99	&88.45±0.0081	&$<0.01$	&1.0  \\
13	&56.32	&58.78±0.0130	&$<0.01$	&$<0.01$  \\
14	&23.94	&25.58±0.0113	&$<0.01$	&$<0.01$  \\
15	&7.82	&8.46±0.0075	&$<0.01$	&$<0.01$  \\
16	&2.40	&2.59±0.0044	&$<0.01$	&$<0.01$  \\
17	&0.62	&0.68±0.0024	&$<0.01$	&$<0.01$  \\
18	&0.12	&0.14±0.0013	&$<0.01$	&$<0.01$  \\
\hline
 \end{tabular}}

\vspace{.5cm}

\subsubsection{CMV indication}

\centerline{
\begin{tabular}{ |c | c | c | c | c |}
 \hline
  $\ell$  & Normal & Lung, $\rho=0.01$& P-value& P-value random \\
    \hline
9	&100.0	&99.99±$(<10^{-5})$	&1.0	&0.4631  \\
10	&99.71	&99.84±0.0004	&$<0.01$	&1.0  \\
11	&94.88	&95.97±0.0017	&$<0.01$	&1.0  \\
12	&74.92	&77.22±0.0031	&$<0.01$	&1.0  \\
13	&43.52	&45.49±0.0035	&$<0.01$	&1.0  \\
14	&18.64	&19.54±0.0029	&$<0.01$	&1.0  \\
15	&6.60	&6.89±0.0019	&$<0.01$	&1.0  \\
16	&2.23	&2.31±0.0011	&$<0.01$	&1.0  \\
17	&0.77	&0.79±0.0007	&$<0.01$	&1.0  \\
18	&0.27	&0.28±0.0004	&$<0.01$	&1.0  \\
\hline
 \end{tabular}}

\vspace{.5cm}

\centerline{
\begin{tabular}{ |c | c | c | c | c |}
 \hline
  $\ell$  & Normal & Bladder, $\rho=0.01$& P-value& P-value random \\
    \hline
9	&100.0	&99.99±$(<10^{-5})$	&1.0	&0.1164  \\
10	&99.71	&99.85±0.0004	&$<0.01$	&0.2682  \\
11	&94.88	&96.06±0.0017	&$<0.01$	&1.0  \\
12	&74.92	&77.45±0.0031	&$<0.01$	&1.0  \\
13	&43.52	&45.67±0.0036	&$<0.01$	&1.0  \\
14	&18.64	&19.63±0.0029	&$<0.01$	&1.0  \\
15	&6.60	&6.93±0.0020	&$<0.01$	&1.0  \\
16	&2.23	&2.33±0.0012	&$<0.01$	&1.0  \\
17	&0.77	&0.79±0.0007	&$<0.01$	&0.9999  \\
18	&0.27	&0.28±0.0004	&$<0.01$	&0.9996  \\
\hline
 \end{tabular}}

\vspace{.5cm}

\centerline{
\begin{tabular}{ |c | c | c | c | c |}
 \hline
  $\ell$  & Normal & Mela, $\rho=0.01$& P-value& P-value random \\
    \hline
9	&100.0	&99.99±$(<10^{-5})$	&1.0	&0.4026  \\
10	&99.71	&99.83±0.0004	&$<0.01$	&1.0  \\
11	&94.88	&95.88±0.0017	&$<0.01$	&1.0  \\
12	&74.92	&76.94±0.0032	&$<0.01$	&1.0  \\
13	&43.52	&45.20±0.0036	&$<0.01$	&1.0  \\
14	&18.64	&19.39±0.0029	&$<0.01$	&1.0  \\
15	&6.60	&6.84±0.0020	&$<0.01$	&1.0  \\
16	&2.23	&2.30±0.0011	&$<0.01$	&1.0  \\
17	&0.77	&0.78±0.0007	&$<0.01$	&1.0  \\
18	&0.27	&0.28±0.0004	&$<0.01$	&0.9999  \\
\hline
 \end{tabular}}

\vspace{.5cm}

\subsubsection{Dengue virus indication}

\centerline{
\begin{tabular}{ |c | c | c | c | c |}
 \hline
  $\ell$  & Normal & Lung, $\rho=0.01$& P-value& P-value random \\
    \hline
9	&100.0	&99.99±$(<10^{-5})$	&1.0	&0.5  \\
10	&100.0	&99.98±0.0002	&1.0	&0.9511  \\
11	&99.30	&99.45±0.0033	&$<0.01$	&1.0  \\
12	&91.65	&92.69±0.0094	&$<0.01$	&0.9996  \\
13	&64.19	&66.20±0.0159	&$<0.01$	&$<0.01$  \\
14	&29.87	&31.21±0.0169	&$<0.01$	&$<0.01$  \\
15	&9.98	&10.52±0.0119	&$<0.01$	&$<0.01$  \\
16	&2.81	&2.99±0.0072	&$<0.01$	&$<0.01$  \\
17	&0.73	&0.78±0.0035	&$<0.01$	&0.0167  \\
18	&0.19	&0.20±0.0019	&$<0.01$	&0.9490  \\
\hline
 \end{tabular}}

\vspace{.5cm}

\centerline{
\begin{tabular}{ |c | c | c | c | c |}
 \hline
  $\ell$  & Normal & Bladder, $\rho=0.01$& P-value& P-value random \\
    \hline
9	&100.0	&99.99±$(<10^{-5})$	&1.0	&0.5  \\
10	&100.0	&99.98±0.0002	&1.0	&0.3175  \\
11	&99.30	&99.47±0.0033	&$<0.01$	&0.9999  \\
12	&91.65	&92.69±0.0094	&$<0.01$	&0.9999  \\
13	&64.19	&66.12±0.0163	&$<0.01$	&$<0.01$  \\
14	&29.87	&31.12±0.0166	&$<0.01$	&$<0.01$  \\
15	&9.98	&10.49±0.0114	&$<0.01$	&$<0.01$  \\
16	&2.81	&2.97±0.0066	&$<0.01$	&$<0.01$  \\
17	&0.73	&0.78±0.0037	&$<0.01$	&0.6529  \\
18	&0.19	&0.20±0.0018	&$<0.01$	&0.9154  \\
\hline
 \end{tabular}}

\vspace{.5cm}

\centerline{
\begin{tabular}{ |c | c | c | c | c |}
 \hline
  $\ell$  & Normal & Mela, $\rho=0.01$& P-value& P-value random \\
    \hline
9	&100.0	&99.99±$(<10^{-5})$	&1.0	&$<0.01$  \\
10	&100.0	&99.98±0.0002	&1.0	&0.9623  \\
11	&99.30	&99.43±0.0032	&$<0.01$	&1.0  \\
12	&91.65	&92.60±0.0094	&$<0.01$	&1.0  \\
13	&64.19	&66.19±0.0168	&$<0.01$	&$<0.01$  \\
14	&29.87	&31.23±0.0163	&$<0.01$	&$<0.01$  \\
15	&9.98	&10.53±0.0114	&$<0.01$	&$<0.01$  \\
16	&2.81	&2.98±0.0068	&$<0.01$	&$<0.01$  \\
17	&0.73	&0.78±0.0035	&$<0.01$	&0.8850  \\
18	&0.19	&0.20±0.0019	&$<0.01$	&0.9918  \\
\hline
 \end{tabular}}

\vspace{.5cm}

\subsubsection{EBV indication}

\centerline{
\begin{tabular}{ |c | c | c | c | c |}
 \hline
  $\ell$  & Normal & Lung, $\rho=0.01$& P-value& P-value random \\
    \hline
9	&100.0	&99.99±$(<10^{-5})$	&1.0	&0.4341  \\
10	&99.94	&99.97±0.0002	&$<0.01$	&0.9999  \\
11	&98.26	&98.69±0.0013	&$<0.01$	&1.0  \\
12	&86.74	&88.26±0.0032	&$<0.01$	&1.0  \\
13	&58.29	&60.10±0.0047	&$<0.01$	&1.0  \\
14	&27.43	&28.38±0.0043	&$<0.01$	&1.0  \\
15	&10.05	&10.34±0.0033	&$<0.01$	&1.0  \\
16	&3.19	&3.28±0.0025	&$<0.01$	&1.0  \\
17	&1.02	&1.04±0.0014	&$<0.01$	&1.0  \\
18	&0.33	&0.33±0.0008	&$<0.01$	&1.0  \\
\hline
 \end{tabular}}

\vspace{.5cm}

\centerline{
\begin{tabular}{ |c | c | c | c | c |}
 \hline
  $\ell$  & Normal & Bladder, $\rho=0.01$& P-value& P-value random \\
    \hline
9	&100.0	&99.99±$(<10^{-5})$	&1.0	&0.1901  \\
10	&99.94	&99.97±0.0002	&$<0.01$	&0.2963  \\
11	&98.26	&98.71±0.0012	&$<0.01$	&1.0  \\
12	&86.74	&88.35±0.0033	&$<0.01$	&1.0  \\
13	&58.29	&60.16±0.0046	&$<0.01$	&1.0  \\
14	&27.43	&28.42±0.0044	&$<0.01$	&1.0  \\
15	&10.05	&10.37±0.0034	&$<0.01$	&1.0  \\
16	&3.19	&3.29±0.0024	&$<0.01$	&1.0  \\
17	&1.02	&1.05±0.0015	&$<0.01$	&1.0  \\
18	&0.33	&0.33±0.0010	&$<0.01$	&1.0  \\
\hline
 \end{tabular}}

\vspace{.5cm}

\centerline{
\begin{tabular}{ |c | c | c | c | c |}
 \hline
  $\ell$  & Normal & Mela, $\rho=0.01$& P-value& P-value random \\
    \hline
9	&100.0	&99.99±$(<10^{-5})$	&1.0	&0.3681  \\
10	&99.94	&99.97±0.0002	&$<0.01$	&1.0  \\
11	&98.26	&98.66±0.0013	&$<0.01$	&1.0  \\
12	&86.74	&88.11±0.0033	&$<0.01$	&1.0  \\
13	&58.29	&59.85±0.0046	&$<0.01$	&1.0  \\
14	&27.43	&28.16±0.0042	&$<0.01$	&1.0  \\
15	&10.05	&10.23±0.0033	&$<0.01$	&1.0  \\
16	&3.19	&3.24±0.0022	&$<0.01$	&1.0  \\
17	&1.02	&1.03±0.0013	&$<0.01$	&1.0  \\
18	&0.33	&0.32±0.0008	&1.0	&1.0  \\
\hline
 \end{tabular}}

\vspace{.5cm}

\subsubsection{HHV indication}

\centerline{
\begin{tabular}{ |c | c | c | c | c |}
 \hline
  $\ell$  & Normal & Lung, $\rho=0.01$& P-value& P-value random \\
    \hline
9	&100.0	&99.99±$(<10^{-5})$	&1.0	&0.7781  \\
10	&99.83	&99.90±0.0004	&$<0.01$	&1.0  \\
11	&95.96	&96.87±0.0020	&$<0.01$	&1.0  \\
12	&79.24	&81.36±0.0050	&$<0.01$	&1.0  \\
13	&49.20	&51.48±0.0060	&$<0.01$	&$<0.01$  \\
14	&21.84	&23.18±0.0050	&$<0.01$	&$<0.01$  \\
15	&7.91	&8.42±0.0031	&$<0.01$	&$<0.01$  \\
16	&2.73	&2.89±0.0021	&$<0.01$	&$<0.01$  \\
17	&1.05	&1.09±0.0015	&$<0.01$	&$<0.01$  \\
18	&0.48	&0.49±0.0017	&$<0.01$	&$<0.01$  \\
\hline
 \end{tabular}}

\vspace{.5cm}

\centerline{
\begin{tabular}{ |c | c | c | c | c |}
 \hline
  $\ell$  & Normal & Bladder, $\rho=0.01$& P-value& P-value random \\
    \hline
9	&100.0	&99.99±$(<10^{-5})$	&1.0	&0.5638  \\
10	&99.83	&99.90±0.0004	&$<0.01$	&0.9999  \\
11	&95.96	&96.93±0.0021	&$<0.01$	&1.0  \\
12	&79.24	&81.48±0.0051	&$<0.01$	&1.0  \\
13	&49.20	&51.50±0.0067	&$<0.01$	&$<0.01$  \\
14	&21.84	&23.15±0.0051	&$<0.01$	&$<0.01$  \\
15	&7.91	&8.41±0.0033	&$<0.01$	&$<0.01$  \\
16	&2.73	&2.89±0.0021	&$<0.01$	&$<0.01$  \\
17	&1.05	&1.10±0.0016	&$<0.01$	&$<0.01$  \\
18	&0.48	&0.49±0.0017	&$<0.01$	&$<0.01$  \\
\hline
 \end{tabular}}

\vspace{.5cm}

\centerline{
\begin{tabular}{ |c | c | c | c | c |}
 \hline
  $\ell$  & Normal & Mela, $\rho=0.01$& P-value& P-value random \\
    \hline
9	&100.0	&99.99±$(<10^{-5})$	&1.0	&0.8672  \\
10	&99.83	&99.89±0.0004	&$<0.01$	&1.0  \\
11	&95.96	&96.79±0.0021	&$<0.01$	&1.0  \\
12	&79.24	&81.22±0.0052	&$<0.01$	&1.0  \\
13	&49.20	&51.50±0.0069	&$<0.01$	&$<0.01$  \\
14	&21.84	&23.26±0.0055	&$<0.01$	&$<0.01$  \\
15	&7.91	&8.48±0.0034	&$<0.01$	&$<0.01$  \\
16	&2.73	&2.91±0.0022	&$<0.01$	&$<0.01$  \\
17	&1.05	&1.11±0.0015	&$<0.01$	&$<0.01$  \\
18	&0.48	&0.49±0.0017	&$<0.01$	&$<0.01$  \\
\hline
 \end{tabular}}

\vspace{.5cm}

\subsubsection{HPV indication}

\centerline{
\begin{tabular}{ |c | c | c | c | c |}
 \hline
  $\ell$  & Normal & Lung, $\rho=0.01$& P-value& P-value random \\
    \hline
9	&100.0	&99.98±$(<10^{-5})$	&1.0	&0.1587  \\
10	&99.97	&99.96±0.0007	&1.0	&0.9984  \\
11	&99.05	&99.23±0.0039	&$<0.01$	&0.9999  \\
12	&90.48	&91.75±0.0114	&$<0.01$	&$<0.01$  \\
13	&61.24	&63.56±0.0183	&$<0.01$	&$<0.01$  \\
14	&28.22	&29.77±0.0175	&$<0.01$	&$<0.01$  \\
15	&9.84	&10.55±0.0124	&$<0.01$	&$<0.01$  \\
16	&3.19	&3.45±0.0076	&$<0.01$	&$<0.01$  \\
17	&1.10	&1.19±0.0046	&$<0.01$	&$<0.01$  \\
18	&0.50	&0.53±0.0025	&$<0.01$	&$<0.01$  \\
\hline
 \end{tabular}}

\vspace{.5cm}

\centerline{
\begin{tabular}{ |c | c | c | c | c |}
 \hline
  $\ell$  & Normal & Bladder, $\rho=0.01$& P-value& P-value random \\
    \hline
9	&100.0	&99.98±$(<10^{-5})$	&1.0	&0.5  \\
10	&99.97	&99.97±0.0007	&1.0	&0.9724  \\
11	&99.05	&99.24±0.0039	&$<0.01$	&0.9999  \\
12	&90.48	&91.71±0.0111	&$<0.01$	&$<0.01$  \\
13	&61.24	&63.34±0.0182	&$<0.01$	&$<0.01$  \\
14	&28.22	&29.65±0.0180	&$<0.01$	&$<0.01$  \\
15	&9.84	&10.48±0.0121	&$<0.01$	&$<0.01$  \\
16	&3.19	&3.43±0.0072	&$<0.01$	&$<0.01$  \\
17	&1.10	&1.17±0.0042	&$<0.01$	&$<0.01$  \\
18	&0.50	&0.52±0.0024	&$<0.01$	&0.5377  \\
\hline
 \end{tabular}}

\vspace{.5cm}

\centerline{
\begin{tabular}{ |c | c | c | c | c |}
 \hline
  $\ell$  & Normal & Mela, $\rho=0.01$& P-value& P-value random \\
    \hline
9	&100.0	&99.98±$(<10^{-5})$	&1.0	&0.1587  \\
10	&99.97	&99.96±0.0007	&1.0	&0.9999  \\
11	&99.05	&99.21±0.0038	&$<0.01$	&1.0  \\
12	&90.48	&91.77±0.0115	&$<0.01$	&$<0.01$  \\
13	&61.24	&63.70±0.0181	&$<0.01$	&$<0.01$  \\
14	&28.22	&29.96±0.0179	&$<0.01$	&$<0.01$  \\
15	&9.84	&10.64±0.0124	&$<0.01$	&$<0.01$  \\
16	&3.19	&3.48±0.0072	&$<0.01$	&$<0.01$  \\
17	&1.10	&1.19±0.0043	&$<0.01$	&$<0.01$  \\
18	&0.50	&0.53±0.0025	&$<0.01$	&$<0.01$  \\
\hline
 \end{tabular}}

\vspace{.5cm}

\subsubsection{Measles virus indication}

\centerline{
\begin{tabular}{ |c | c | c | c | c |}
 \hline
  $\ell$  & Normal & Lung, $\rho=0.01$& P-value& P-value random \\
    \hline
9	&100.0	&99.99±$(<10^{-5})$	&1.0	&0.5  \\
10	&99.97	&99.98±0.0004	&$<0.01$	&0.9998  \\
11	&98.23	&98.70±0.0037	&$<0.01$	&1.0  \\
12	&86.15	&87.87±0.0094	&$<0.01$	&0.9934  \\
13	&54.27	&56.70±0.0140	&$<0.01$	&$<0.01$  \\
14	&22.64	&23.98±0.0127	&$<0.01$	&$<0.01$  \\
15	&7.29	&7.73±0.0084	&$<0.01$	&$<0.01$  \\
16	&2.19	&2.29±0.0049	&$<0.01$	&0.5766  \\
17	&0.68	&0.69±0.0027	&$<0.01$	&0.3373  \\
18	&0.19	&0.19±0.0013	&0.5424	&0.0179  \\
\hline
 \end{tabular}}

\vspace{.5cm}

\centerline{
\begin{tabular}{ |c | c | c | c | c |}
 \hline
  $\ell$  & Normal & Bladder, $\rho=0.01$& P-value& P-value random \\
    \hline
9	&100.0	&99.99±$(<10^{-5})$	&1.0	&0.5  \\
10	&99.97	&99.98±0.0004	&$<0.01$	&0.9992  \\
11	&98.23	&98.71±0.0037	&$<0.01$	&1.0  \\
12	&86.15	&87.88±0.0091	&$<0.01$	&0.3703  \\
13	&54.27	&56.57±0.0138	&$<0.01$	&$<0.01$  \\
14	&22.64	&23.86±0.0124	&$<0.01$	&$<0.01$  \\
15	&7.29	&7.69±0.0078	&$<0.01$	&0.5972  \\
16	&2.19	&2.28±0.0049	&$<0.01$	&0.9990  \\
17	&0.68	&0.69±0.0026	&$<0.01$	&0.7406  \\
18	&0.19	&0.19±0.0014	&0.5426	&0.0220  \\
\hline
 \end{tabular}}

\vspace{.5cm}

\centerline{
\begin{tabular}{ |c | c | c | c | c |}
 \hline
  $\ell$  & Normal & Mela, $\rho=0.01$& P-value& P-value random \\
    \hline
9	&100.0	&99.99±$(<10^{-5})$	&1.0	&1.0  \\
10	&99.97	&99.98±0.0004	&$<0.01$	&0.9999  \\
11	&98.23	&98.65±0.0038	&$<0.01$	&1.0  \\
12	&86.15	&87.81±0.0096	&$<0.01$	&1.0  \\
13	&54.27	&56.61±0.0131	&$<0.01$	&$<0.01$  \\
14	&22.64	&23.92±0.0121	&$<0.01$	&$<0.01$  \\
15	&7.29	&7.71±0.0082	&$<0.01$	&0.0134  \\
16	&2.19	&2.27±0.0048	&$<0.01$	&0.9999  \\
17	&0.68	&0.68±0.0026	&0.0137	&0.9609  \\
18	&0.19	&0.19±0.0014	&0.9999	&0.9017  \\
\hline
 \end{tabular}}

\vspace{.5cm}

\subsubsection{Yellow fever virus indication}

\centerline{
\begin{tabular}{ |c | c | c | c | c |}
 \hline
  $\ell$  & Normal & Lung, $\rho=0.01$& P-value& P-value random \\
    \hline
9	&100.0	&99.99±$(<10^{-5})$	&1.0	&0.5  \\
10	&100.0	&99.98±0.0002	&1.0	&0.6664  \\
11	&99.05	&99.30±0.0034	&$<0.01$	&1.0  \\
12	&92.10	&93.10±0.0094	&$<0.01$	&1.0  \\
13	&65.28	&67.19±0.0169	&$<0.01$	&$<0.01$  \\
14	&30.54	&31.81±0.0176	&$<0.01$	&$<0.01$  \\
15	&10.81	&11.26±0.0119	&$<0.01$	&$<0.01$  \\
16	&3.15	&3.32±0.0070	&$<0.01$	&0.3854  \\
17	&0.87	&0.94±0.0039	&$<0.01$	&$<0.01$  \\
18	&0.29	&0.31±0.0022	&$<0.01$	&$<0.01$  \\
\hline
 \end{tabular}}

\vspace{.5cm}

\centerline{
\begin{tabular}{ |c | c | c | c | c |}
 \hline
  $\ell$  & Normal & Bladder, $\rho=0.01$& P-value& P-value random \\
    \hline
9	&100.0	&99.99±$(<10^{-5})$	&1.0	&0.5  \\
10	&100.0	&99.98±0.0003	&1.0	&0.9609  \\
11	&99.05	&99.30±0.0034	&$<0.01$	&0.9999  \\
12	&92.10	&93.11±0.0096	&$<0.01$	&0.9999  \\
13	&65.28	&67.09±0.0173	&$<0.01$	&0.9999  \\
14	&30.54	&31.75±0.0164	&$<0.01$	&0.0112  \\
15	&10.81	&11.25±0.0116	&$<0.01$	&0.0579  \\
16	&3.15	&3.32±0.0068	&$<0.01$	&0.1432  \\
17	&0.87	&0.94±0.0038	&$<0.01$	&$<0.01$  \\
18	&0.29	&0.32±0.0023	&$<0.01$	&$<0.01$  \\
\hline
 \end{tabular}}

\vspace{.5cm}

\centerline{
\begin{tabular}{ |c | c | c | c | c |}
 \hline
  $\ell$  & Normal & Mela, $\rho=0.01$& P-value& P-value random \\
    \hline
9	&100.0	&99.99±$(<10^{-5})$	&1.0	&$<0.01$  \\
10	&100.0	&99.98±0.0003	&1.0	&0.9888  \\
11	&99.05	&99.26±0.0036	&$<0.01$	&1.0  \\
12	&92.10	&93.01±0.0090	&$<0.01$	&1.0  \\
13	&65.28	&67.03±0.0170	&$<0.01$	&1.0  \\
14	&30.54	&31.77±0.0166	&$<0.01$	&$<0.01$  \\
15	&10.81	&11.26±0.0121	&$<0.01$	&$<0.01$  \\
16	&3.15	&3.33±0.0073	&$<0.01$	&$<0.01$  \\
17	&0.87	&0.95±0.0040	&$<0.01$	&$<0.01$  \\
18	&0.29	&0.32±0.0023	&$<0.01$	&$<0.01$  \\
\hline
 \end{tabular}}

\vspace{.5cm}

\subsection{Error probability data for Fig.~\ref{fig:plot2}}\label{cb}
To compute ${\mathbb{E}}|\{i: \tilde{a}_i\ne a_i\}|$ in \eqref{per} we proceeded as follows. We have 
\begin{align}
{\mathbb{E}}|\{i: \tilde{a}_i\ne a_i\}|=\sum_i \mathbb{P}(\tilde{a}_i\ne a_i)
\end{align}
where the summation ranges over amino acid positions. Let us compute $\mathbb{P}(\tilde{a}_1\ne a_1)$---for the other terms we proceed in the same way. 
{
Observe that ${a}_1$ is a function of the first three nucleotides $y_1,y_2,y_3$ of the normal exome $y$.}
 To emphasize this, let us write ${a}_1$ as $a_1({y}_1,{y}_2,{y}_3)$. Similarly, $\tilde{a}_1$ is a function of the first three nucleotides $\tilde{y}_1,\tilde{y}_2,\tilde{y}_3$ of the cancer genome $\tilde{y}$ and we write it as $\tilde{a}_1(\tilde{y}_1,\tilde{y}_2,\tilde{y}_3)$. Therefore, we have

\begin{align}
\mathbb{P}(\tilde{a}_1\ne a_1)=\mathop{\mathop{\sum}_{(\tilde{y}_1,\tilde{y}_2,\tilde{y}_3):}}_{\tilde{a}_1(\tilde{y}_1,\tilde{y}_2,\tilde{y}_3)\ne a_1({y}_1,{y}_2,{y}_3)}\prod_{j=1}^3 \mathbb{P}_{c,\rho}(\tilde{y}_j|y_j)
\end{align} 
where $ \mathbb{P}_{c,\rho}(\tilde{y}_j|y_j)$ is the cancer channel defined in the  Appendix~\ref{appB}.

%
%
%

\section*{Acknowledgements}
The authors would like to thank Ka Kit Lam for early discussions on this work.

\end{document}